 \documentclass[twocolumn]{aastex6}

\fullcollaborationName{The Friends of AASTeX Collaboration}
\begin{document}


\title{Angular momentum in bipolar outflows: dynamical evolutionary model}



\author{J.A. L\'opez-V\'azquez\altaffilmark{1}, J. Cant\'o\altaffilmark{2}, and S. Lizano\altaffilmark{1}}


\altaffiltext{1}{Instituto de Radioastronom\'ia y Astrof\'isica, Universidad Nacional Aut\'onoma de M\'exico, Apartado Postal 3-72, 58089 Morelia, Michoac\'an, M\'exico}
\altaffiltext{2}{Instituto de Astronom\'ia, Universidad Nacional Aut\'onoma de M\'exico, Apartado Postal 70-264, 04510, CDMX, M\'exico}


\begin{abstract}

We model molecular outflows produced by the time dependent interaction between a stellar wind and a rotating cloud envelope in gravitational collapse, studied by Ulrich. We consider spherical and anisotropic stellar winds. 
We assume that the bipolar outflow is a thin shocked shell, with axial symmetry around the cloud rotation axis and obtain the mass and momentum fluxes into the shell. We solve numerically a set of partial differential equations in space and time, and obtain the shape of the shell, the mass surface density, the velocity field, and the angular momentum of the material in the shell. We find that there is a critical value of the ratio between the wind and the accretion flow momentum rates $\beta$ that allows the shell to expand. As expected, the elongation of the shells increase with  the stellar wind anisotropy.
In our models, the rotation velocity of the shell is the order to 0.1 - 0.2 km s$^{-1}$, a factor of 5-10 lower than the values measured in several sources. We compare our models with those of Wilkin and Stahler for early evolutionary times and find that our shells have the same
sizes at the pole, although we use different boundary conditions at the equator.
\end{abstract}



\section{Introduction}
\label{sec:introduction}

The study of the molecular outflows and protostellar jets is fundamental to understand the star formation process. 
Molecular outflows probably limit the mass of the star-disk system (e.g., \citealt{Shu_1993}) and can induce changes in the chemical composition of their host cloud (e.g., \citealt{Bachiller_1996}) since they are  a mixture of entrained material from the cloud and the outflowing stellar wind (e.g., \citealt{Snell_1980}).

The magneto-centrifugal mechanism \citep{Blandford_1982} is considered the principal candidate for producing the jets of young stars (see reviews by \citealt{Konigl_2000} and \citealt{Shu_2000}). In this mechanism the magnetic field, anchored to the star-disk system, is responsable for accelerating the jet. However, it is still under debate where these magnetic fields are anchored to the disk: it could be at a narrow region at the truncation radius $R_x$ of the disk by a stellar magnetosphere (X-winds, e.g., \citealt{Shu_1994}) or at a wider range of radii  (disk winds, e.g.,
\citealt{Pudritz_1983}).
Magnetohydrodynamic models predict that the material ejected from the disk has a toroidal angular momentum component related to the rotation at the disk foot point. Therefore, the observed rotational velocity of the jet can give information about its origin on the disk \citep{Anderson_2003}. For example, \citet{Lee_2009} and \citet{Lee_2017} found that the protostellar jets HH 211 and HH 212, respectively, are ejected from very small radii, consistent with the X-wind model.

\begin{table*}[!t]
 \centering
 \setlength\tabcolsep{2.0pt}
\caption{Observational parameters of the molecular outflows with rotation. 
}
\begin{tabular}{c c c c c c c}
 \hline
 \hline
\textbf{Source}& \textbf{Observations}&$M_*$&$\Delta v_\varphi$&$\Delta r$&z$_\mathrm{cut}$\\
 & &(M$_\odot$)&(km s$^{-1}$)&(AU)&(AU) \\
\hline
CB26 & HCO$^{+}$(1-0) and $^{13}$CO (2-1) &0.5  & 1.1 &100& 560\\
Ori-S6 & SO (6$_5$-5$_4$) and $^{12}$CO (2-1)&2.0&2.0&1000& 1200\\
HH 797 & $^{12}$CO (2-1) & 1.0 &2.0&1000& 7500\\
DG Tau B & $^{12}$CO (2-1) & 0.5 & 1.0 & 150 & 450\\
Orion Source I &  Si$^{18}$O and H$_2$O & 8.7 & 5.0 & 80 & 150\\
HH 30 & $^{12}$CO (2-1) and $^{13}$CO (2-1) & 0.45 & 0.4 & 150 & 200\\
NGC 1333 IRAS 4C & CCH & 0.18& 0.4 & 470 & 700\\
  \hline
  \end{tabular}
\footnote{The first column shows the outflow name, the second column gives the molecular lines observed, the third column indicates the mass of the central star $M_*$, the fourth column shows the velocity difference across the outflow lobe (rotation velocity) $\Delta v_\phi$, the fifth and sixth columns are the distance to the flow axis $\Delta r$, and the height above the disk $z_\mathrm{cut}$, respectively.}
  \label{tab:outflows}  
\end{table*}

Molecular outflows have been explained as driven by fast stellar winds or as actual disk winds. 
In the former case,  molecular outflows are produced when a fast stellar wind collides with the parent cloud accelerating and entraining cloud material
(e.g., see reviews by \citealt{Arce_2007} and \citealt{Bally_2016}). In the latter case, the molecular outflow is ejected directly from the accretion disk (e.g., \citealt{Pudritz_1986}).

In recent years, rotation has been observed  in a few molecular outflows, which are almost on the plane of the sky.
\footnote{As an alternative to the interpretation of a velocity difference as rotation, \citet{Decolle_2016} showed that such differences can also be due to asymmetric shocks produced in the interaction of the stellar wind with the environment or by asymmetries in the ejection velocity of the disk-star system.}
 These sources are: CB 26 \citep{Launhardt_2009}, Ori-S6 \citep{Zapata_2010}, HH 797 \citep{Pech_2012}, DG Tau B \citep{Zapata_2015}, Orion Source I \citep{Hirota_2017}, HH 30 \citep{Louvet_2018}, and NGC 1333 IRAS 4C \citep{Zhang_2018}. Table \ref{tab:outflows}  presents a summary of their characteristics. 

From the observed rotation velocities, assuming that the outflows are disk winds,  different authors obtained a disk launching radii between 10-50 AU (e.g., \citealt{Launhardt_2009}; \citealt{Pech_2012}). Nevertheless, \citet{Zapata_2015}, showed that magneto-centrifugal and photoevaporated disk winds do not have enough linear or angular momentum to account for the observed linear momentum and angular momentum rates in the molecular outflow of DG Tau B. They found that the observed rates are larger by a factor of 100, because the disk winds are not very massive. They pointed out that to account for the large masses of the observed molecular outflows they must be mainly entrained material from the parent cloud.

Several authors have modeled  the molecular outflow as a wind-driven shell formed by the interaction between a radial stellar wind and the ambient cloud (e.g., \citealt{Shu_1991}; \citealt{Matzner_1999}; \citealt{Canto_2006}). The ambient cloud can also be an accreting envelope.
For example, \citet{Mendoza_2004} described the hydrodynamical interaction between a rotating accretion flow and a spherically symmetric stellar wind. However, they did not consider neither the gravitational pull from the central star nor the centrifugal terms in the momentum equation. Later, \citet{Wilkin_2003} took 
into account these effect but considered only the early evolution of the outflow.

Here we present a time dependent model of the interaction between a rotating accretion flow and a fast radial stellar wind. 
The molecular outflow is a thin shell driven by the fast stellar wind, that gains mass from both the stellar wind and the accretion flow. In this model we consider the gravitational pull of the central star and the centrifugal terms in the momentum equations. Also, we follow the evolution of the shell from the stellar surface up to large distances from the central star. We consider the molecular outflows produced by 
both isotropic and axisymmetric stellar winds with a polar angle dependance.

This paper is organized in the following way: in section \ref{sec:general_formulation} we show the equations of the dynamic evolution of the shell. Section \ref{sec:winds} presents the description of the accretion flow and the stellar wind. The method of solution is presented in section \ref{sec:solution}, where we show the non dimensional equations and boundary conditions. In this section we also find semi analytic solutions for expansions around both the pole and the equator. Section \ref{sec:results} presents the results for different stellar winds. In section \ref{sec:discussion} we discuss our results. Finally, the conclusions are presented in section \ref{sec:conclusions}.

\section{General formulation}
\label{sec:general_formulation}

We assume that the molecular outflow is formed by the supersonic collision between a stellar wind and an accretion flow. This collision leads to the formation of both an inner and an outer shock front. We assume that the cooling behind these shocks is relatively efficient because the shock velocities are expected to be less than $<$ 100 km s$^{-1}$ \citep{Hartigan_1987}. Thus, the region between the shocks is described by a cold and thin shell. Within the shell, two fluids with a different density and velocity come into contact producing internal shearing layers which are subject to the Kelvin-Helmholtz instability, quickly leading to a turbulent mixing. Here we assume that the mixing is so efficient that one may describe the shell as a single fluid (e.g., \citealt{Wilkin_2003}).

The evolution of the shell is governed  by the fluxes of mass and momentum from the stellar wind and the accretion flow, by the gravitational influence of the central star, and by the centrifugal effects.

\begin{figure*}[t]
\centering
\includegraphics[scale=0.4]{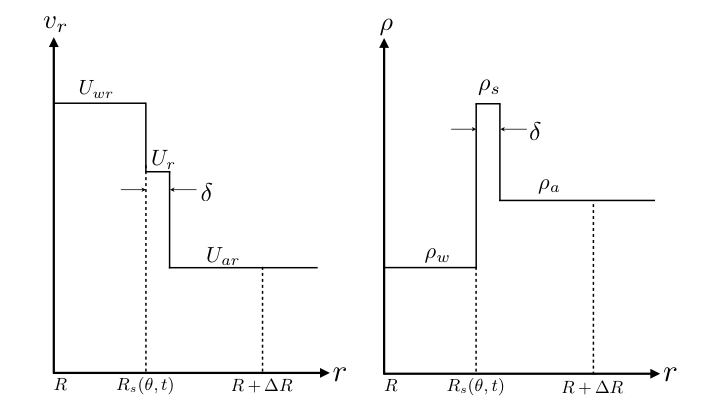}
\caption{Schematic diagram of the model of a thin shell of thickness $\delta$. Left panel: the radial velocity structure around the shell. Right panel: the mass volume density structure around the shell. $U_r$ and $\rho_s$ are the values of the radial velocity and the mass density of the shell and $U_{wr}$, $U_{ar}$, $\rho_{w}$, and $\rho_{a}$ are the values of the fluid velocities and the mass density near the shell which is at $R_s(\theta,t)$. 
We assume that $\delta\ll\Delta R\ll R\simeq R_s(\theta,t)$.}
\label{fig:integration_schemes}
\end{figure*}

\subsection{Shell equations}
\label{sub:mathematical}
To derive the shell equations, we use spherical coordinates $r$, $\theta$, and $\phi$ for the radial, the polar, and the azimuthal coordinates, respectively. The coordinate system is centered on the star and we assume axial symmetry.
The shell has a radius $R_s$, a mass surface density $\sigma$, and velocity components $U_r$, $U_\theta$, and $U_\phi$. All
of these functions depend on $\theta$ and $t$, although, for simplicity, we will omit these dependences.

We assume that the accretion and wind flows are axisymmetric, and that they vary in a timescale much longer than the shell evolution time. Therefore, their 
properties depend only on the coordinates  $r$ and $\theta$.
The accretion flow has a mass volume density $\rho_a$ and velocity components $U_{ar}$, $U_{a\theta}$, and $U_{a\phi}$. The stellar wind has a mass volume density $\rho_w$ and velocity components $U_{wr}$, $U_{w\theta}$, and $U_{w\phi}$.
Figure \ref{fig:integration_schemes} shows the outflow model where a thin shell is formed by the interaction of the stellar wind and the accretion flow.

In Appendix \ref{app:derivation}, we show the derivation of the equations of the shell evolution in a general form. In order to write these equations in more compact form, we define the mass flux
\begin{eqnarray}
P_m=R_s^2\sin\theta\sigma,
\label{eq:pm}
\end{eqnarray}

\noindent and the momentum fluxes
\begin{eqnarray}
P_{r}&=&R_s^2\sin\theta\sigma U_{r}\equiv P_m U_{r},\nonumber \\
P_{\theta}&=&R_s^2\sin\theta\sigma U_{\theta}\equiv P_m U_{\theta},\nonumber \\
P_{\phi}&=&R_s^2\sin\theta\sigma U_{\phi}\equiv P_m U_{\phi}.
\label{eq:prtp}
\end{eqnarray}

\noindent Also, we consider that the stellar wind has only a radial velocity component ($v_w=U_{wr}$).

Then, the continuity equation (eq. [\ref{eq:appmass}]) can be written in terms of the mass and momentum fluxes as
\begin{eqnarray}
&&\frac{\partial P_m}{\partial t}+\frac{\partial}{\partial\theta}\left(\frac{P_\theta}{R_s}\right)= \nonumber \\
& & R_s^2\sin\theta\left[ \rho_a\left(\frac{P_r}{P_m}-U_{ar}\right)-\rho_w\left(\frac{P_r}{P_m}-v_w \right)\right],
\label{eq:pmass}
\end{eqnarray}

\noindent where the RHS shows the contribution to the shell mass from the stellar wind and the accretion flow.

The equation of the momentum in radial direction (eq. [\ref{eq:appmomr}]) is given by
\begin{eqnarray}
&&\frac{\partial P_r}{\partial t}+\frac{\partial}{\partial\theta}\left( \frac{P_r P_\theta}{R_sP_m}\right)-\frac{P_\theta^2+P_\phi^2}{R_sP_m}+\frac{GM_* P_m}{R_s^2}=R_s^2\sin\theta\nonumber \\
&\times& \left[\rho_aU_{ar}\left(\frac{P_r}{P_m}-U_{ar}\right)-\rho_w v_w \left(\frac{P_r}{P_m}-v_w\right)\right],
\label{eq:pmr}
\end{eqnarray}
\noindent where, $G$ is the gravitational constant and $M_*$ is the stellar mass. The third term in the LHS comes from the centrifugal effect, and the last term is due to the weight of the shell. The RHS has the contribution from the stellar wind and the accretion flow.

The momenta in the $\theta$ and the azimuthal directions (eqns. [\ref{eq:appmomtheta}] and [\ref{eq:appmomphi}]) in terms of the mass and momentum fluxes, respectively, can be written as
\begin{eqnarray}
\frac{\partial P_\theta}{\partial t}&+&\frac{\partial}{\partial\theta}\left( \frac{P_\theta^2}{R_sP_m}\right)+\frac{P_rP_\theta-P_\phi^2\cot\theta}{R_sP_m}= \nonumber \\
&&R_s^2\sin\theta\rho_aU_{a\theta}\left(\frac{P_r}{P_m}-U_{ar}\right),
\label{eq:pmtheta}
\end{eqnarray}

\begin{eqnarray}
\frac{\partial P_\phi}{\partial t}&+&\frac{\partial}{\partial\theta}\left( \frac{P_\phi P_\theta}{R_sP_m}\right)+\frac{P_\phi\left(P_r+P_\theta\cot\theta\right)}{R_sP_m}=\nonumber \\
& &R_s^2\sin\theta\rho_aU_{a\phi}\left(\frac{P_r}{P_m}-U_{ar}\right).
\label{eq:pmphi}
\end{eqnarray}
In these two equations, the last terms on the LHS are due to the centrifugal effect on the shell, while in the RHS only the accretion flow contributes to the momentum fluxes in these directions.

Finally, the evolution of the shell radius can be written as
\begin{eqnarray}
\frac{\partial R_s}{\partial t}=\frac{P_r}{P_m}-\frac{1}{R_s}\frac{P_\theta}{P_m}\frac{\partial R_s}{\partial\theta},
\label{eq:prshell}
\end{eqnarray}
where the first term in the RHS corresponds to the radial velocity and the second term is the contribution of the tangential motion along the shell.

To solve these equations for the evolution of the shell radius $R_s(\theta,t)$, one needs to specify the properties of the accretion flow and the stellar wind.
This model allows an accretion flow and a stellar wind with a general velocity field. The only constrain is that they
have to be axisymmetric. We note that our formulation of the equations is different from \citet{Wilkin_2003}. In our model, the vectors are expressed in spherical coordinates, while in the model of \citet{Wilkin_2003} the vectors are decomposed in directions orthogonal and parallel to the shell.

\section{Accretion flow and stellar wind}
\label{sec:winds}
In this section we will apply our model to flows with  specific properties  in order to solve 
for the shell evolution.

\subsection{The accretion flow}
\label{sub:accretion}
The accretion flow is given by the gravitational collapse of a rotating cloud described by \citet{Ulrich_1976}. In this model, the fluid particles have a uniform rotation rate at large distances and fall to the center conserving the specific angular momentum $j$. The collapse is assumed to be pressureless and thus, the orbits are ballistic. The latter assumption is true when the flow is supersonic, and heating by radiation and viscosity effects are negligible. The collapse of the gas reaches a centrifugal barrier at $R_\mathrm{cen}=j^2/GM_*$.

In terms of the non dimensional radial variable
\begin{eqnarray}
\zeta\equiv \frac{R_\mathrm{cen}}{r},
\label{eq:zeta1}
\end{eqnarray}

\noindent and the polar angle $\theta$, the velocity field and the density profile of the accretion flow are given by
\begin{eqnarray}
U_{ar}=-v_0\zeta^{1/2}\left(1+\frac{\cos\theta}{\cos\theta_0} \right)^{1/2},
\label{eq:uar}
\end{eqnarray}
\begin{eqnarray}
U_{a\theta}=v_0\zeta^{1/2}\left(\frac{\cos\theta_0-\cos\theta}{\sin\theta} \right)\left(1+\frac{\cos\theta}{\cos\theta_0} \right)^{1/2},
\label{eq:uat}
\end{eqnarray}
\begin{eqnarray}
U_{a\phi}=-v_0\zeta^{1/2}\frac{\sin\theta_0}{\sin\theta}\left(1-\frac{\cos\theta}{\cos\theta_0} \right)^{1/2},
\label{eq:uap}
\end{eqnarray}
\noindent and
\begin{eqnarray}
\rho_a=-\frac{\dot{M}_{a}\zeta^2}{4\pi R_{cen}^2U_{ar}}[1+2\zeta P_2(\cos\theta_0)]^{-1},
\label{eq:rhoa}
\end{eqnarray}

\noindent where $\theta_0$ is the initial polar angle of the orbit of the fluid element at the beginning of the collapse towards the center, $v_0$ is the free fall velocity
\begin{eqnarray}
v_0=\left(\frac{GM_*}{R_\mathrm{cen}} \right)^{1/2},
\label{eq:v0}
\end{eqnarray}

\noindent $\dot{M}_a$ is the mass accretion rate, and the Legendre polynomial is $P_2(\cos\theta_0)=\frac{1}{2}\left(3\cos\theta_0^2-1\right)$. The angle $\theta_0$ is given implicitly in terms of variables $\theta$ and $\zeta$ by
\begin{eqnarray}
\zeta=\frac{\cos\theta_0-\cos\theta}{\sin^2\theta_0\cos\theta_0}.
\label{eq:zeta}
\end{eqnarray}

\noindent Eq. (13) of \citet{Mendoza_2004} gives an explicit solution of this equation.

Note that eq. (\ref{eq:uap}) for the azimuthal velocity differes in sign with that given by \citet{Ulrich_1976}. This only means that we consider the accretion flow to rotate with a negative angular momentum, as in Figure 1 of \citet{Mendoza_2004}.

\subsection{The stellar wind}
\label{sub:stellarwind}
We assume an anisotropic stellar wind with a mass loss rate $\dot{M}_w$ and only a radial velocity component $U_{wr}=v_w$, 
assumed to be constant. The density is given by
\begin{eqnarray}
\rho_w=\frac{\dot{M}_w}{4\pi r^2 v_w}f(\theta),
\label{eq:rhow}
\end{eqnarray}

\noindent where $f(\theta)$ is anisotropy function given by
\begin{eqnarray}
f(\theta)=\frac{A+B\cos^{2n}\theta}{A+B/(2n+1)}.
\label{eq:fanisotropic}
\end{eqnarray}

\noindent The constants are $A\geq 0$ and $B\geq 0$ and $n$ is an integer. For $B=0$ or $n=0$ one recovers an isotropic stellar wind, while for $B>0$ and $n>0$, the density profile is anisotropic. This function is normalized such that the integral of the mass flux around the star recovers the total mass loss rate, $\dot{M}_w=2\pi\int^{\pi}_0\rho_wv_w r^2\sin\theta d\theta$.

\section{Solution of the equations}
\label{sec:solution}
\subsection{Non dimensional equations}
\label{sub:nondimensional}
To solve the equations we define the following non dimensional variables: the non dimensional radius
\begin{eqnarray}
r_s=\frac{R_s}{R_\mathrm{cen}},
\label{eq:rsad}
\end{eqnarray}

\noindent the non dimensional time
\begin{eqnarray}
\tau=\frac{v_0}{R_\mathrm{cen}}t,
\label{eq:tau}
\end{eqnarray}

\noindent the non dimensional mass flux
\begin{eqnarray}
p_m=\frac{4\pi v_0}{\dot{M}_a R_\mathrm{cen}}P_m,
\label{eq:pmad}
\end{eqnarray} 

\noindent and the non dimensional momentum fluxes
\begin{eqnarray}
p_{r}=\frac{4\pi}{\dot{M}_a R_\mathrm{cen}}P_{r},\\
p_{\theta}=\frac{4\pi}{\dot{M}_a R_\mathrm{cen}}P_{\theta}, \\
p_{\phi}=\frac{4\pi}{\dot{M}_a R_\mathrm{cen}}P_{\phi}.
\label{eq:pmad}
\end{eqnarray}

The non dimensional velocities of the accretion flow and the stellar wind are
\begin{eqnarray}
u_{ar}=\frac{U_{ar}}{v_0},\\
u_{a\theta}=\frac{U_{a\theta}}{v_0}, \\
u_{a\phi}=\frac{U_{a\phi}}{v_0},
\label{eq:velaccretion}
\end{eqnarray}

\noindent and
\begin{eqnarray}
u_{wr}=\frac{v_w}{v_0}.
\label{eq:velstellar}
\end{eqnarray}

Also, we define the ratio of the wind mass loss rate and the mass accretion rate
\begin{eqnarray}
\alpha=\frac{\dot{M}_w}{\dot{M}_a},
\label{eq:alpha}
\end{eqnarray}

\noindent and the ratio between the stellar wind and the accretion flow momentum rates
\begin{eqnarray}
\beta=\frac{\dot{M}_w v_w}{\dot{M}_a v_0}\equiv\alpha u_{wr}.
\label{eq:beta}
\end{eqnarray}

\noindent Finally, the non dimensional densities of the accretion flow and the stellar wind are given by
\begin{eqnarray}
\rho_a^\prime=\frac{4\pi R_{cen}^2 v_0}{\dot{M}_a}\rho_a\equiv-\frac{\zeta^2}{u_{ar}}\left[1+2\zeta P_2(\cos\theta_0)\right]^{-1},
\label{eq:rhoandim}
\end{eqnarray}

\noindent and
\begin{eqnarray}
\rho_w^\prime=\frac{4\pi R_{cen}^2v_0}{\dot{M}_a}\rho_w\equiv\frac{\alpha}{r_s^2u_{wr}}f(\theta).
\label{eq:rhowndim}
\end{eqnarray}

In terms of the new variables, eqns. (\ref{eq:pmass}) - (\ref{eq:prshell}) can be written as
\begin{eqnarray}
&&\frac{\partial p_m}{\partial\tau}+\frac{\partial}{\partial\theta}\left(\frac{p_\theta}{r_s}\right)=\sin\theta\times \nonumber \\
&&\left[\frac{ \left(u_{ar}-\frac{p_r}{p_m}\right)}{u_{ar}\left[1+2\zeta P_2(\cos\theta_0)\right]}-\alpha f(\theta)\left(\frac{\alpha}{\beta}\frac{p_r}{p_m}-1\right)\right],
\label{eq:tildepm}
\end{eqnarray}

\begin{eqnarray}
&&\frac{\partial p_r}{\partial\tau}+\frac{\partial}{\partial\theta}\left(\frac{p_r p_\theta}{r_s p_m}\right)-\frac{p_\theta^2+p_\phi^2}{r_s p_m}+\frac{p_m}{r_s^2}=\sin\theta\times\nonumber \\
&&\left[\frac{ \left(u_{ar}-\frac{p_r}{p_m}\right)}{1+2\zeta P_2(\cos\theta_0)}-\beta f(\theta)\left(\frac{\alpha}{\beta}\frac{p_r}{p_m}-1\right)\right],
\label{eq:tildepr}
\end{eqnarray}

\begin{eqnarray}
&&\frac{\partial p_\theta}{\partial\tau}+\frac{\partial}{\partial\theta}\left(\frac{p_\theta^2  }{r_s p_m}\right)+\frac{p_r p_\theta-p_\phi^2\cot\theta}{r_s p_m}=\nonumber \\
&&\frac{\sin\theta}{1+2\zeta P_2(\cos\theta_0)}\left(\frac{u_{a\theta}}{u_{ar}} \right) \left(u_{ar}-\frac{p_r}{p_m}\right),
\label{eq:tildeptheta}
\end{eqnarray}

\begin{eqnarray}
&&\frac{\partial p_\phi}{\partial\tau}+\frac{\partial}{\partial\theta}\left(\frac{p_\phi p_\theta}{r_s p_m}\right)+\frac{p_\phi \left(p_r+p_\theta\cot\theta\right)}{r_s p_m}=\nonumber \\
&&\frac{\sin\theta}{1+2\zeta P_2(\cos\theta_0)}\left(\frac{u_{a\phi}}{u_{ar}} \right) \left(u_{ar}-\frac{p_r}{p_m} \right),
\label{eq:tildepphi}
\end{eqnarray}

\begin{eqnarray}
\frac{\partial r_s}{\partial\tau}=\frac{p_r}{p_m}-\frac{1}{r_s}\frac{p_\theta}{p_m}\frac{\partial r_s}{\partial\theta},
\label{eq:trs}
\end{eqnarray}

\noindent where $f(\theta)$ is defined in eq. (\ref{eq:fanisotropic}).

These equations need initial conditions to advance in time, and boundary conditions (BCs) at the pole ($\theta=0$) and at the equator ($\theta=\pi/2$).

\subsection{Boundary conditions}
\label{sub:boundary}
In the next section we expand the variables in powers of $\theta$ and obtain equations for their time evolution at the pole and at the equator. These solutions provide BCs for the partial differential equations (\ref{eq:tildepm}) - (\ref{eq:trs}).

\subsubsection{Expansions around the pole}
\label{sub:pole}
We expand in power series the mass and momentum fluxes as well as the radius of the shell. The equations are expanded to second order in $\theta$ for $\theta\ll1$, such that the variables are given by
\begin{eqnarray}
p_m\approx b_{m1}\theta,
\label{eq:polepm}
\end{eqnarray}

\begin{eqnarray}
p_r\approx b_{r1}\theta,
\label{eq:polepr}
\end{eqnarray}

\begin{eqnarray}
p_\theta\approx b_{\theta2}\theta^2,
\label{eq:poleptheta}
\end{eqnarray}

\begin{eqnarray}
p_\phi\approx b_{\phi2}\theta^2,
\label{eq:polepphi}
\end{eqnarray}

\noindent and
\begin{eqnarray}
r_s\approx r_{s0},
\label{eq:polers}
\end{eqnarray}

\noindent where the coefficients $b_{m1}$, $b_{r1}$, $b_{\theta2}$, $b_{\phi2}$, and $r_{s0}$ are functions of the non dimensional time ($\tau$) given by the solution of the set of differential equations described in Appendix \ref{app:pole}.

\subsubsection{Expansions around the equator}
\label{sub:equator}
In the equator the density of the accretion flow at centrifugal radius diverges. If the shell evolves in this direction, eventually, it is going to find a barrier of infinite density. At that point the shell will stagnate at $R_{cen}$ or collapse back to stellar surface.

We expand the variables around the equator as
\begin{eqnarray}
p_m\approx q_{m0}+q_{m1} \Theta, 
\label{eq:pmeq}
\end{eqnarray}

\begin{eqnarray}
p_r\approx q_{r0}+q_{r1}\Theta, 
\label{eq:preq}
\end{eqnarray}

\begin{eqnarray}
p_\theta\approx q_{\theta0}+q_{\theta1} \Theta,
\label{eq:pteq}
\end{eqnarray}

\begin{eqnarray}
p_\phi\approx q_{\phi0}+q_{\phi1} \Theta,
\label{eq:ppeq}
\end{eqnarray}

\noindent and
\begin{eqnarray}
r_s\approx q_{rs0}+q_{rs1} \Theta, 
\label{eq:radeq}
\end{eqnarray}
where $\Theta= \left(\frac{\pi}{2}-\eta\right)-\theta \ll 1$, and 
the angle $\eta$ defines a physical boundary in the equatorial region (e.g., a disk).

The coefficients $q_{m0}$, $q_{m1}$, $q_{r0}$, $q_{r1}$, $q_{\theta0}$, $q_{\theta1}$, $q_{\phi0}$, $q_{\phi1}$, $q_{rs0}$, and $q_{rs1}$ are functions of the non dimensional time ($\tau$) given by the solution of a set of differential equations described in Appendix \ref{app:equator}.

\section{Results}
\label{sec:results}

\begin{figure}[h!]
\centering
\includegraphics[scale=0.55]{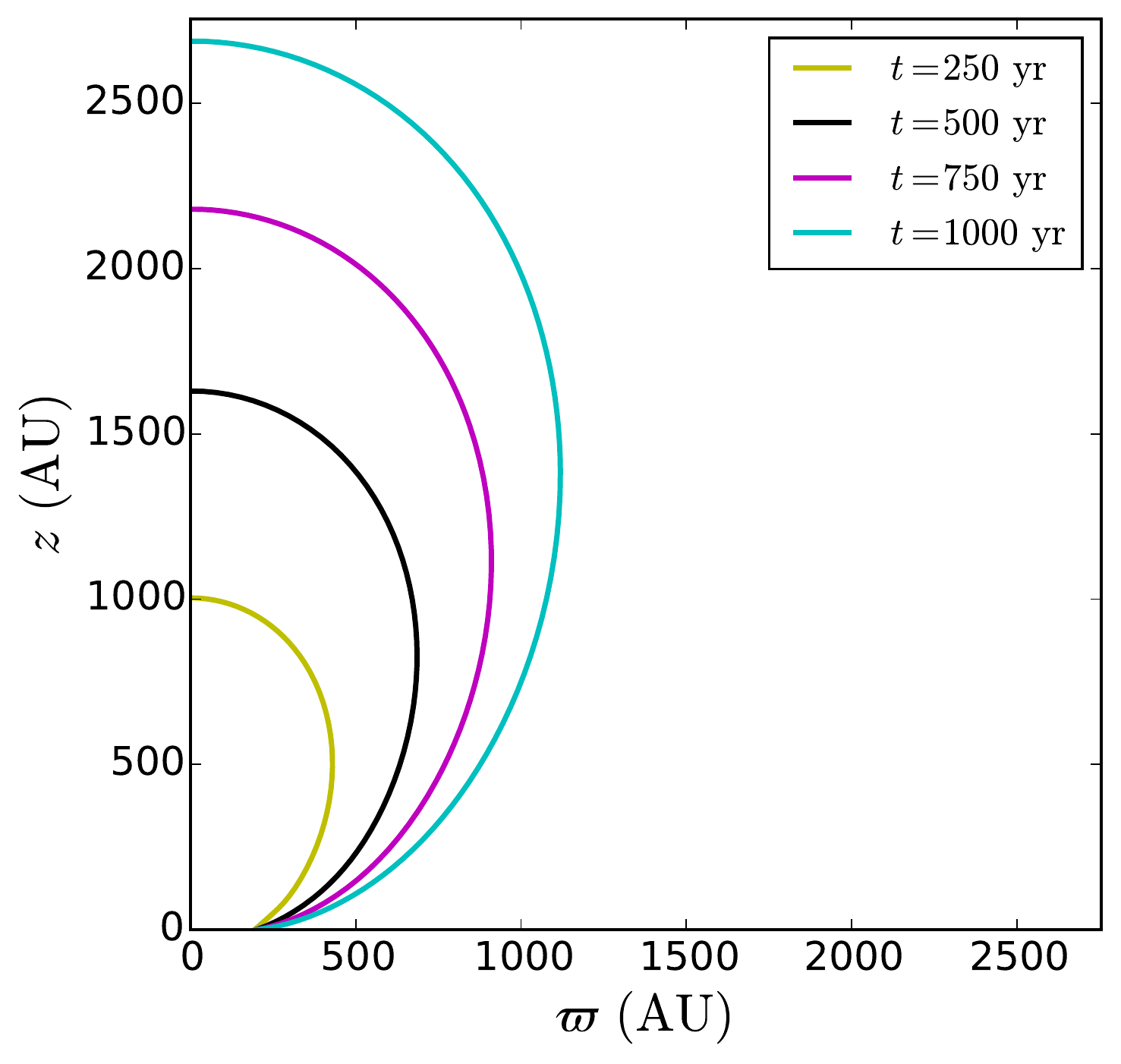}
\caption{Shape of the shell for different times for the parameters $\alpha=0.1$, $\beta=21$, $r_{s0}(0)=10^{-4}$, $A=1$, $B=20$, and $n=2$.}
\label{fig:eshell}
\end{figure}

\begin{figure*}[!t]
\raggedleft
\includegraphics[scale=0.425]{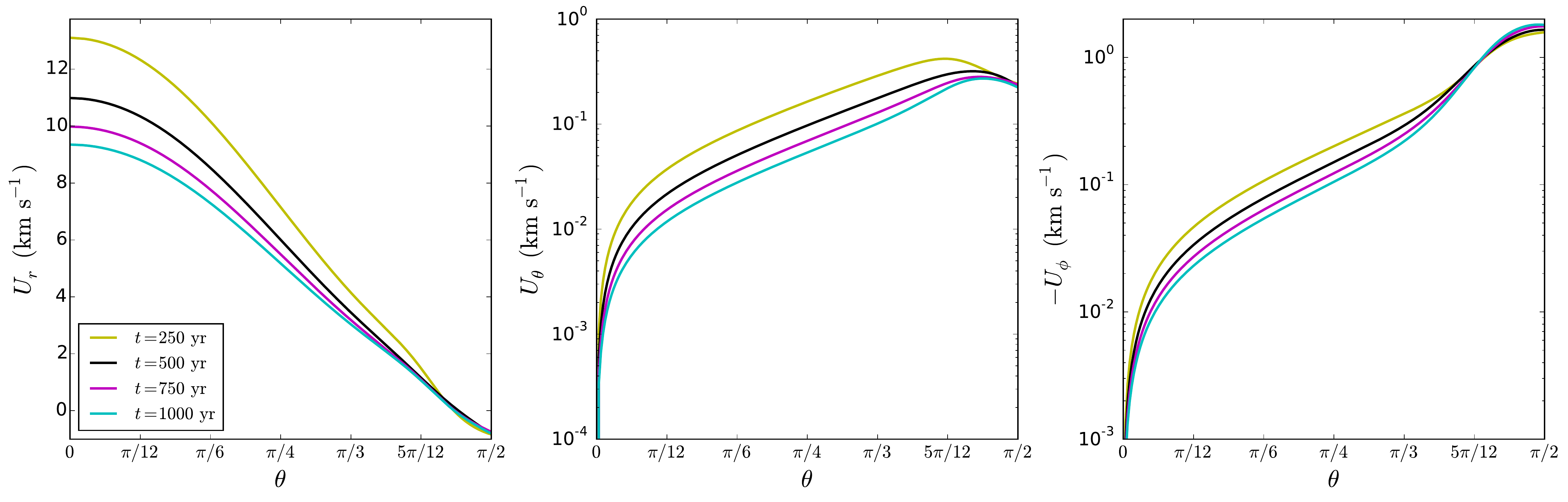}
\caption{Velocity field of the shell as a function of $\theta$ for same parameters $\alpha$, $\beta$, $r_{s0}(0)$, $A$, $B$, and $n$ as figure \ref{fig:eshell}. Left panel: the radial velocity. Middle panel: the $\theta$-velocity. Right panel: the azimuthal velocity.}
\label{fig:shellvelocity}
\end{figure*}

Eqns. (\ref{eq:tildepm}) - (\ref{eq:trs}) describe the evolution of the shell. The non dimensional equations are solved numerically. We assume that initially, the shell is spherical and massless, with a radius close to the stellar surface. We assume an initial non dimensional radius of $r_{s0}(0)\simeq R_*/R_{cen}\simeq10^{-4}$. We assume that the ratio of the wind mass loss rate and the mass accretion rate is $\alpha=0.1$, a typical value for molecular outflows (e.g., see figure 14 of \citealt{Ellerbroek_2013}). Also, we assume $\beta=21$\footnote{The stellar wind and the free fall velocities correspond to parameters of the central star of the molecular outflow CB 26 \citep{Launhardt_2009}. These parameters are a stellar mass $\mathrm{M}_*=0.5\mathrm{M}_\odot$ and a centrifugal radius of $R_{cen}=200$ AU \citep{Launhardt_2001}. 
We also assumed a radius $R_*=2$ R$_\odot$. The stellar wind velocity is taken as the escape velocity of the central star. 
With these assumptions, the stellar wind velocity of $309$ km s$^{-1}$ and free fall velocity of $v_0=1.5$ km s$^{-1}$. }. The integration is done from $t=0$ to  $t=1000$ yr.

In this section we study the shell evolution for different stellar wind models. As an example, consider the molecular outflow produced by an anisotropic stellar wind with $A=1$, $B=20$, and $n=2$.  Figure \ref{fig:eshell} shows the shape of the shell $R_s(\theta, t)$ for different times from $t=250$ yr to $t=1000$ yr. 
The shells are elongated along the cloud rotational axis. We define the shell collimation as the ratio
\begin{eqnarray}
C=\frac{R_s(0,t)}{\varpi_{\rm max}(t)},
\label{eq:collimation}
\end{eqnarray}

\noindent where  $R_s(0,t)$ is the shell radius at the pole, and $\varpi_{\rm max}$  is the maximum width 
of the shell. This ratio measures the shell elongation. During the shell evolution, this model has a collimation 
$C \sim 2.5$, similar to the observed outflows CB 26 \citep{Launhardt_2009} and DG Tau B \citep{Zapata_2015}.  

Figure \ref{fig:shellvelocity} shows the radial, the $\theta$, and the azimuthal velocities as functions of $\theta$ for this model, for different times. 
The velocities of the shell are obtained from eqns. (\ref{eq:prtp}).
The left panel shows the radial velocity of the shell. This velocity decreases with time, i.e., the shell is slowing down. It also decreases with the angle such that at the equator, $\theta=\pi/2$ this velocity tends zero, because the shell finds a barrier at $R_{\rm cen}$
where the density is infinite. The middle panel shows the $\theta$-velocity of the shell.
This velocity decreases with the time, but increases with the polar angle $\theta$, due to the material that slides from the pole to the equator; this material feeds the accretion disk $U_\theta>0$. The right panel shows the azimuthal velocity. This rotation velocity decreases with the time and increases with angle: at the equator the rotation velocity is maximum because the orbits of the accretion flow  that lands at this point have the largest angular momentum with respect to the pole.

\begin{figure}[h!]
\centering
\includegraphics[scale=0.5]{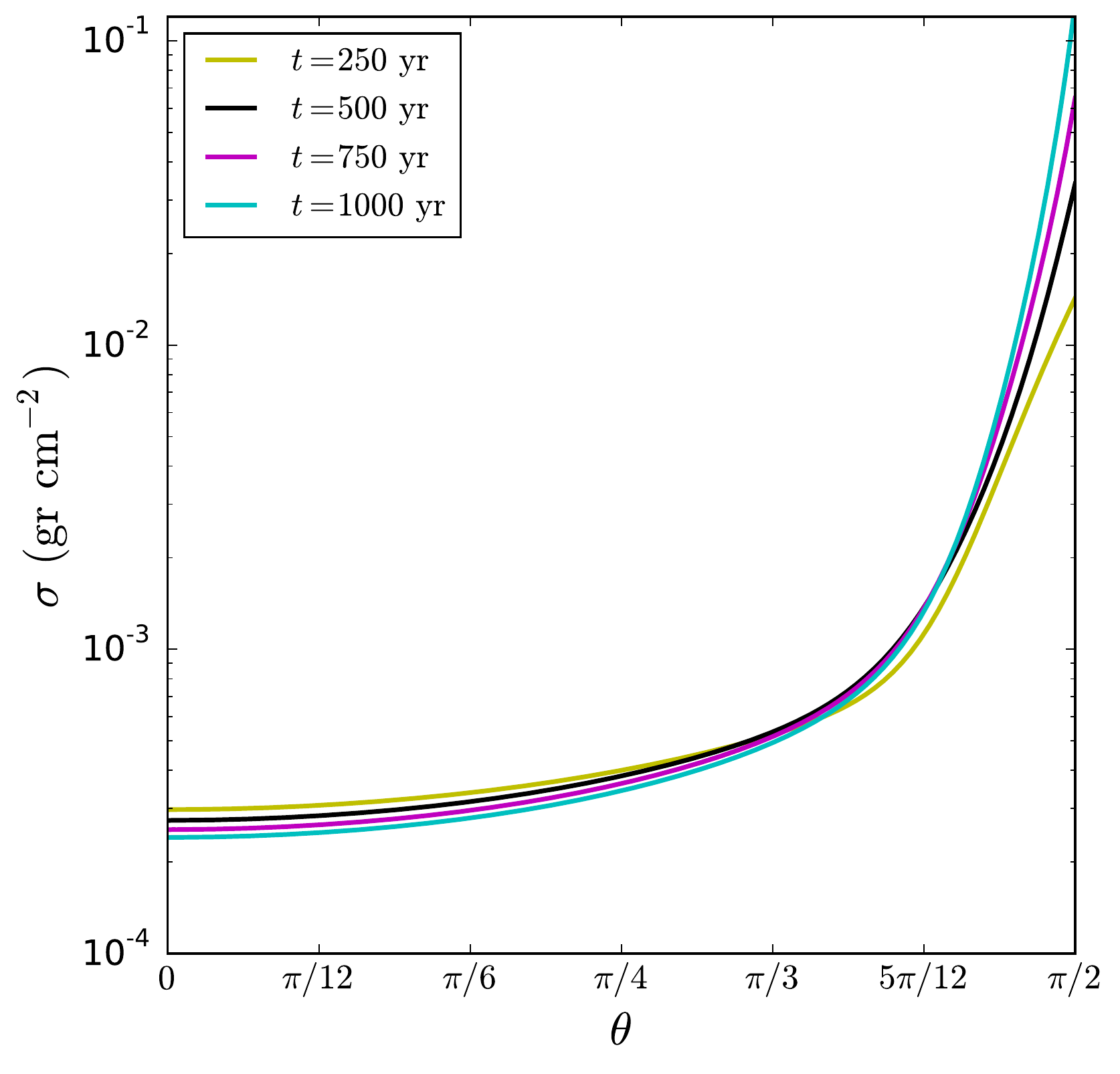}
\caption{The mass surface density of the shell as a function of $\theta$ for same parameters $\alpha$, $\beta$, $r_{s0}(0)$, $A$, $B$, and $n$ as figure \ref{fig:eshell}.}
\label{fig:sigmashell}
\end{figure}

The mass surface density of the shell along the radial direction, obtained from eq. (\ref{eq:pm}), is plotted in Figure \ref{fig:sigmashell}. In this figure, we observe that, for angles close to the pole, the surface density decreases with time, while for angles close to the equator, the surface density increases with time.

\begin{figure}[h!]
\centering
\includegraphics[scale=0.5]{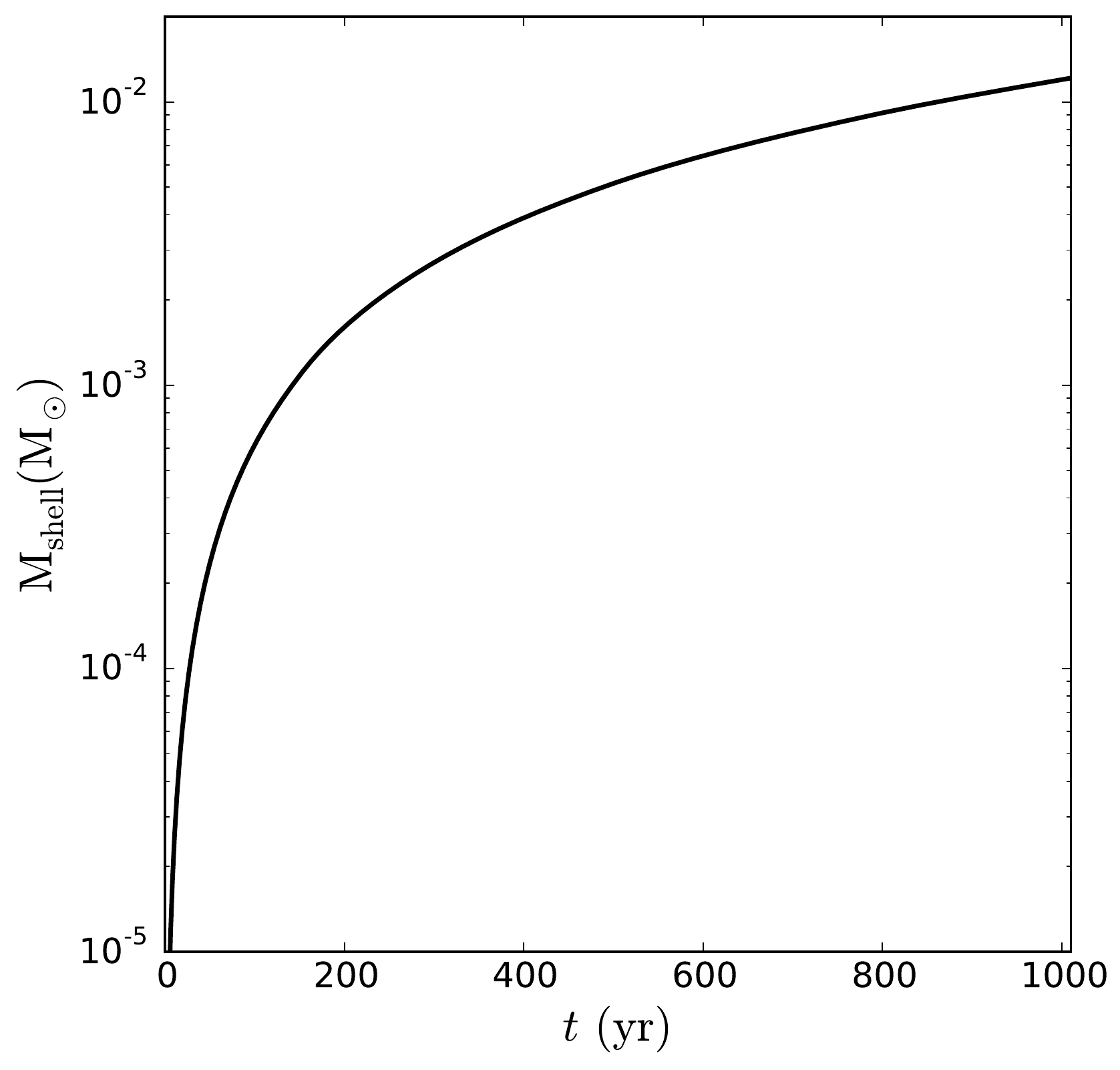}
\caption{The total mass of the shell as a function of time for same parameters $\alpha$, $\beta$, $r_{s0}(0)$, $A$, $B$, and $n$ as figure \ref{fig:eshell}.}
\label{fig:massshell}
\end{figure}

The total mass of the shell is given by
\begin{eqnarray}
M_\mathrm{shell}(t)= 2 \int_0^{\pi/2} \sigma d A = 4\pi\int_0^{\pi/2}P_m d\theta,
\label{eq:totalmass}
\end{eqnarray}

\noindent where $d A = 2 \pi R^2 \sin\theta d\theta$, and $P_m$ defined in eq. (\ref{eq:pm}). Figure \ref{fig:massshell} shows that the mass increases with time.

\begin{figure}[h!]
\centering
\includegraphics[scale=0.5]{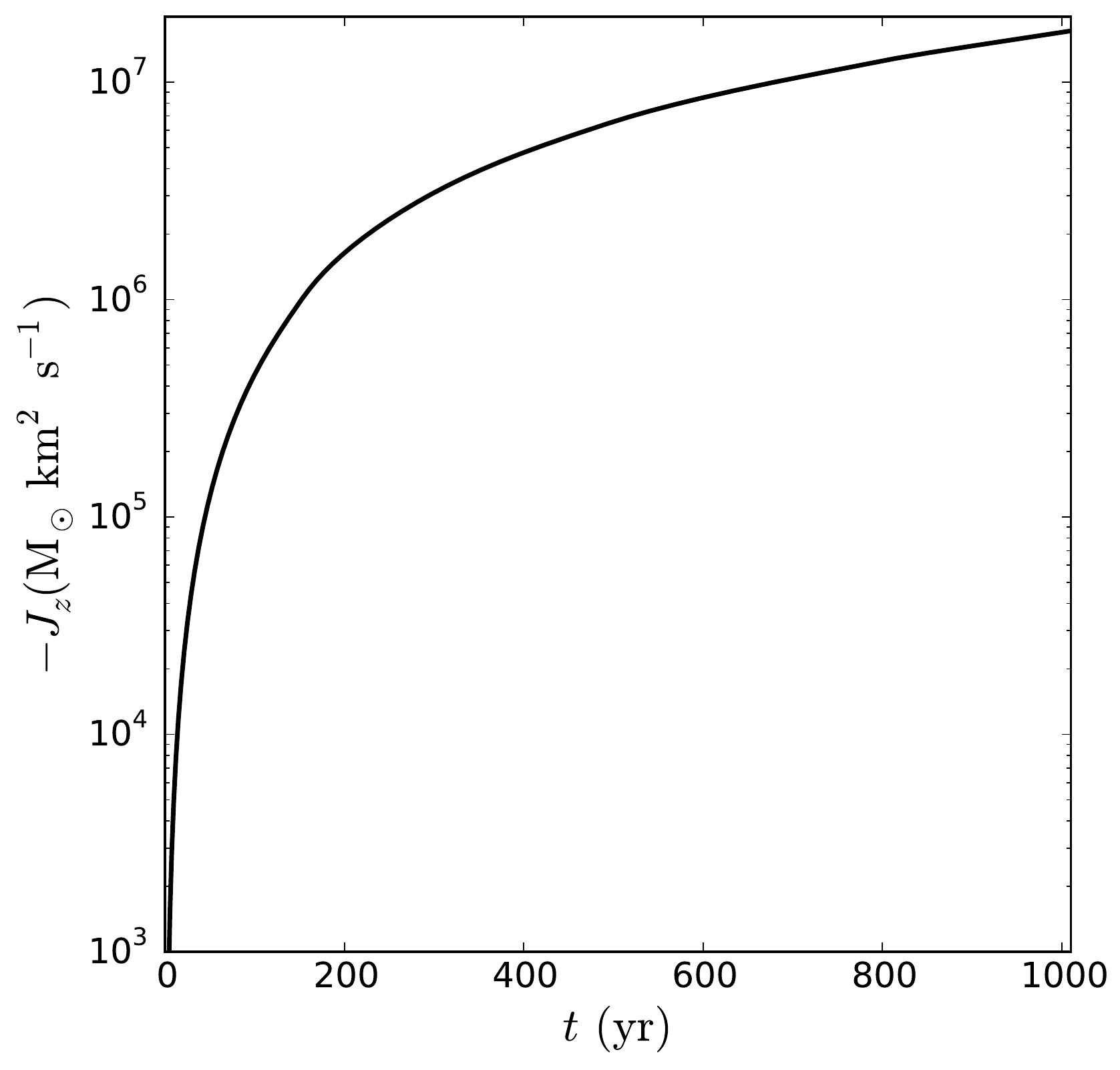}
\caption{The total angular momentum of the shell as a function of time for same parameters $\alpha$, $\beta$, $r_{s0}(0)$, $A$, $B$, and $n$ as figure \ref{fig:eshell}.}
\label{fig:jzshell}
\end{figure}

The specific angular momentum of the shell in $z-$direction is
\begin{eqnarray}
{j}_z(\theta,t)=U_\phi R_s\sin\theta,
\label{eq:angz}
\end{eqnarray}

\noindent and the total angular momentum is
\begin{eqnarray}
J_z(t)=2 \int_0^{\pi/2} \sigma j_z d A = 4 \pi \int_0^{\pi/2} P_m j_z  d\theta.
\label{eq:jzshell}
\end{eqnarray}

\noindent Figure \ref{fig:jzshell} shows the total angular momentum, which, increases with time.

\begin{figure*}[h!]
\centering
\includegraphics[scale=0.525]{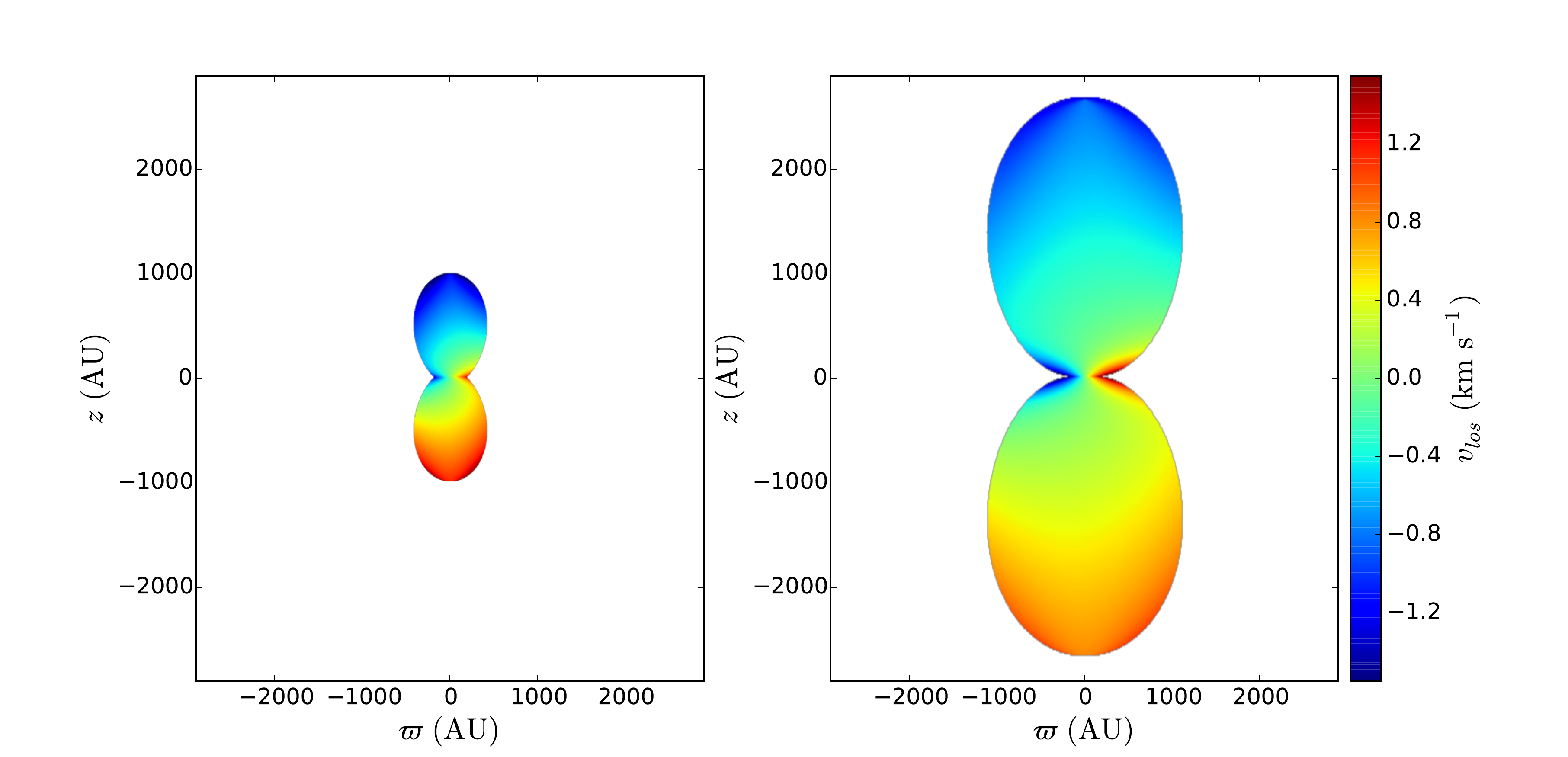}
\caption{Velocity of the line of sight for different times and an inclination angle $i=5^\circ$. Left panel: velocity of the line of sight for a time $t=250$ yr. Right panel: velocity of the line of sight for a time $t=1000$ yr. These plots were made for the same parameters $\alpha$, $\beta$, $r_{s0}(0)$, $A$, $B$, and $n$ as figure \ref{fig:eshell}.}
\label{fig:ulosshell}
\end{figure*}

\begin{figure*}[h!]
\raggedleft
\includegraphics[scale=0.425]{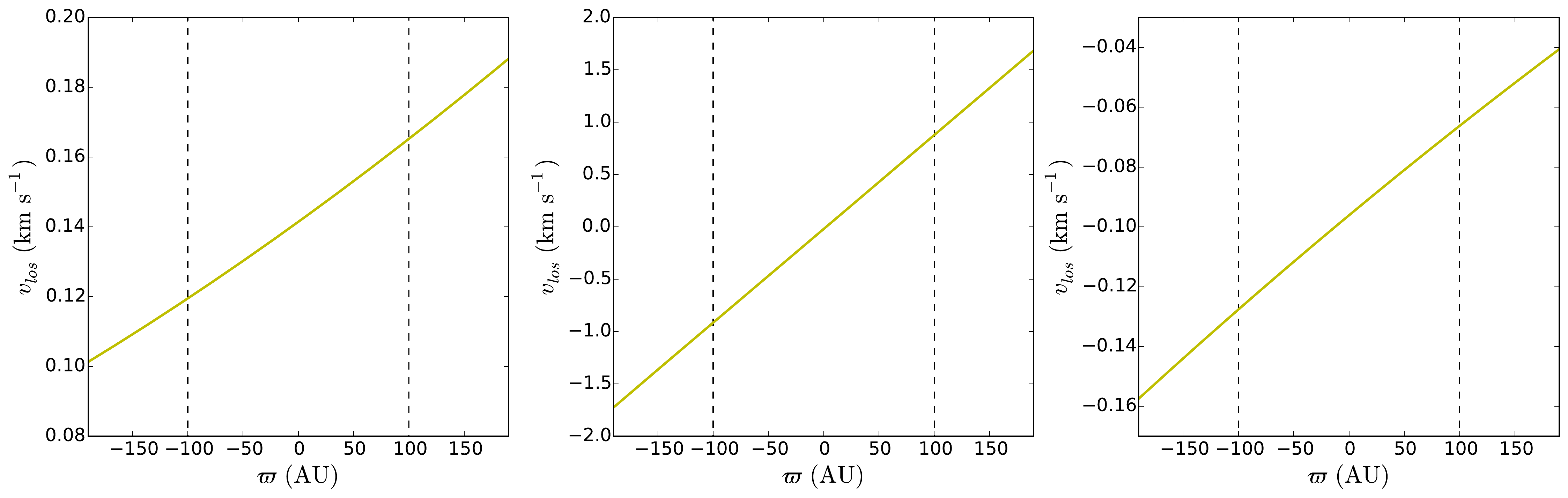}
\caption{Position-velocity diagrams with perpendicular cuts to the cloud's rotational axis for different heights from the disk and an inclination angle $i=5^\circ$. Left panel: velocity of the line of sight for $z_{cut}=-560$ AU. Middle panel: velocity of the line of sight for the disk midplane. Right panel: velocity of the line of sight for $z_{cut}=420$ AU. These plots were made for same parameters $\alpha$, $\beta$, $r_{s0}(0)$, $A$, $B$, and $n$ as figure \ref{fig:eshell}.}
\label{fig:zcut}
\end{figure*}

\begin{figure*}[h!]
\centering
\includegraphics[scale=0.55]{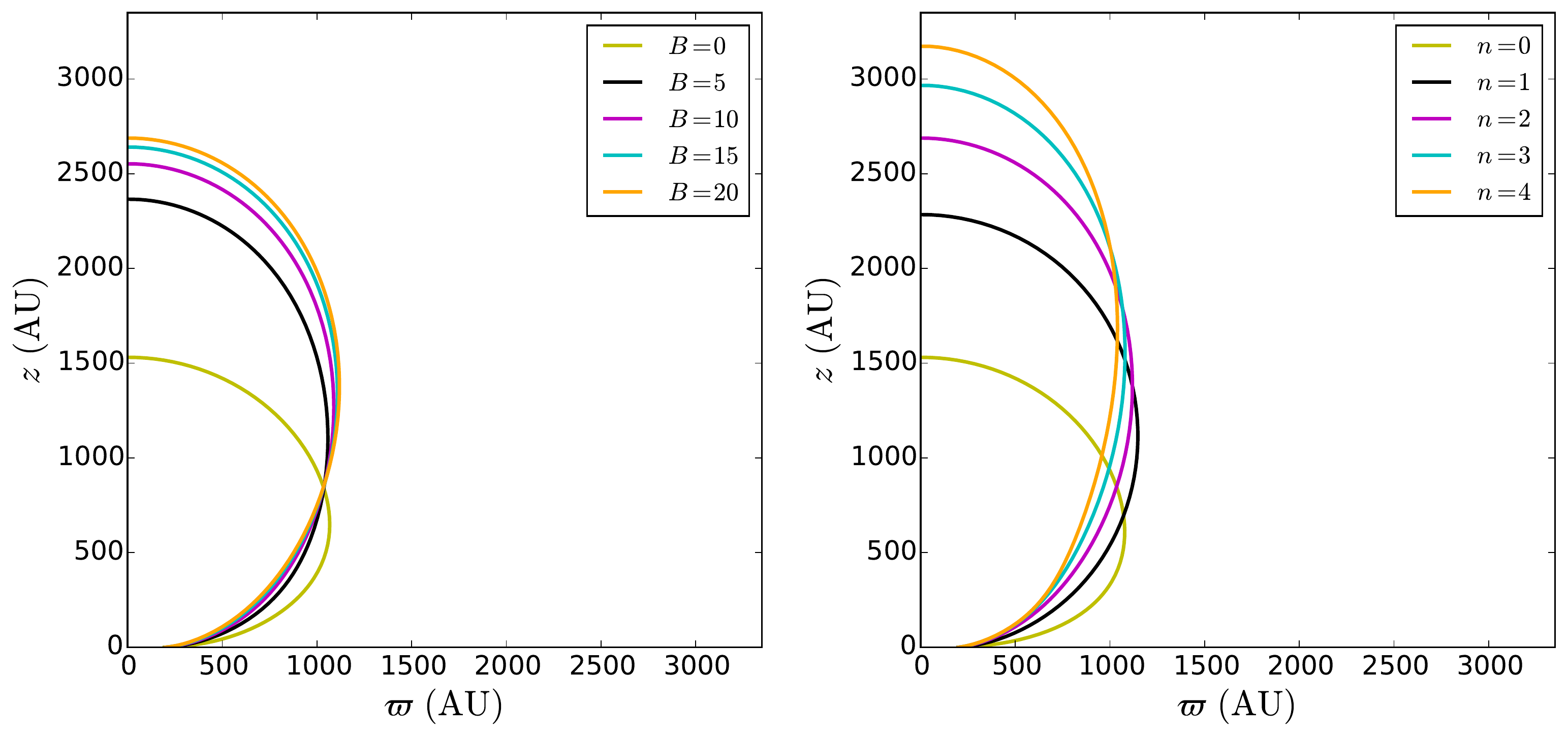}
\caption{Shape of the shell for different models at $t=1000$ yr and the same parameters $\alpha$, $\beta$, and $r_{s0}(0)$ as figure \ref{fig:eshell}. Left panel: stellar wind model with $A=1$, $n=2$, and different values of the anisotropy parameter $B=0$, 5, 10, 15, and 20. Right panel:  anisotropic stellar wind with $A=1$, $B=20$, and different exponents $n=0$, 1, 2, 3, and 4.}
\label{fig:shellanisotropy}
\end{figure*}

\begin{figure}[h!]
\centering
\includegraphics[scale=0.5]{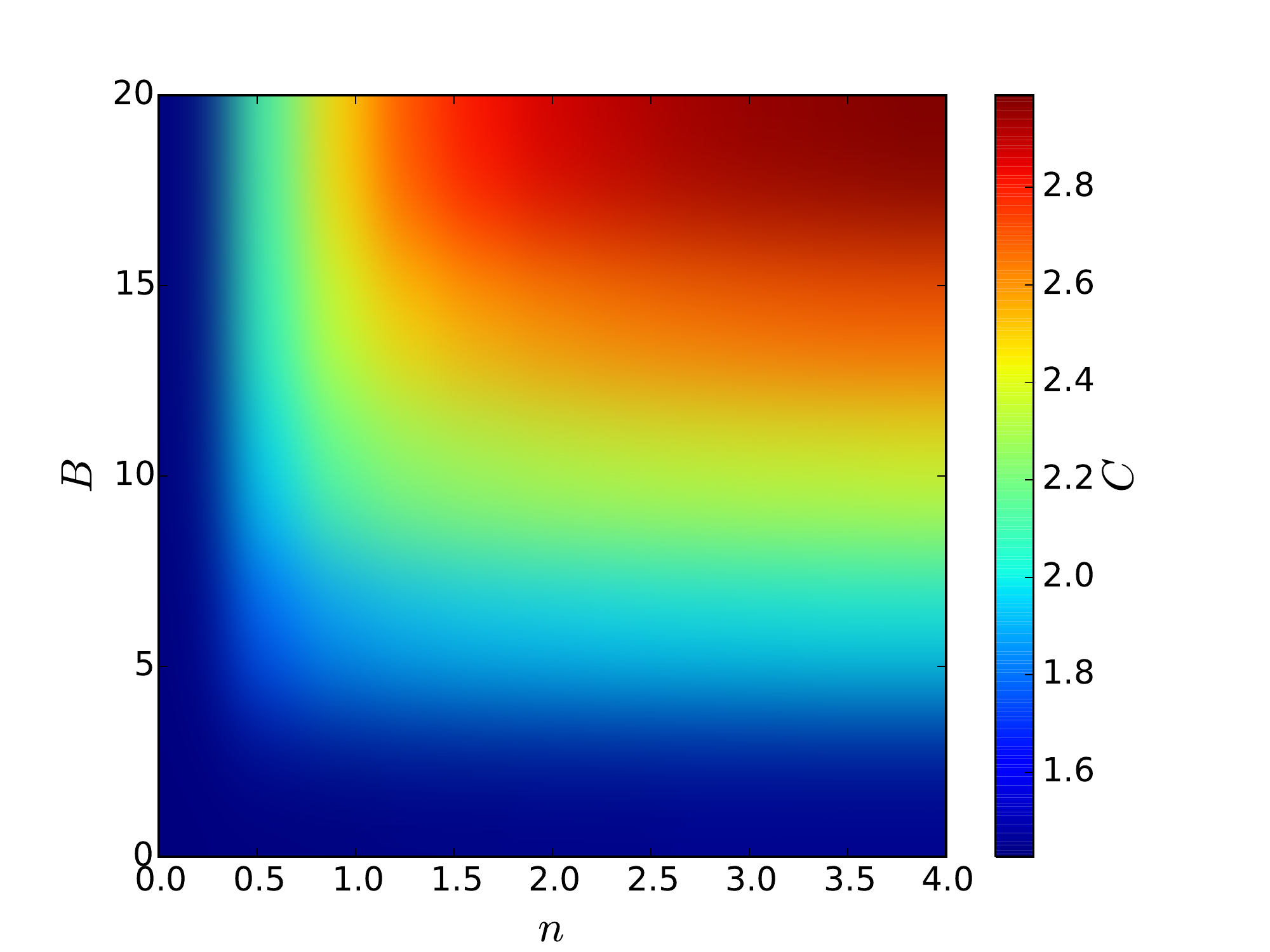}
\caption{Collimation $C$ as a function of $B$ and $n$ for $A=1$ and the same parameters $\alpha$, $\beta$, and $r_{s0}(0)$ as figure \ref{fig:eshell}.}
\label{fig:collimation}
\end{figure}

In order to compare the outflow model with observations, we projected the velocity field of the shell along the line of sight for an inclination angle $i=5^\circ$ with respect to the plane of the sky. The velocity of the line of sight $v_{los}$ is shown in Figure \ref{fig:ulosshell}. The left and right panels show the velocity at 250 yr and 1000 yr, respectively. For an inclination angle larger than $0^\circ$, the velocity along the line of sight is a combination of the radial and the $\theta$ velocities. Figure \ref{fig:zcut} shows cuts at different heights $z_{cut}$ of the map at 250 yr. The left panel shows a position-velocity diagram for $z_{cut}=-560$ AU. The middle panel and right panel have cuts at $z_{cut}=0$ and $z_{cut}=420$, respectively.

Now we consider the effect of degree of anisotropy of the stellar wind on the shape of the shell. Figure \ref{fig:shellanisotropy} shows the shape of the shell  for parameters $\alpha=0.1$, $\beta=21$, $r_{s0}(0)=10^{-4}$, and a stellar wind model with $A=1$, a time of $t=1000$ yr, for different values of the anisotropy parameters $B$ and $n$. The left panel shows the shape of the shells for $n=2$, one can see that as $B$ increases, the shell becomes more elongated. The right panel shows the shape shell for $B=20$, as 
the exponent $n$ increases, the shell are more elongated.

Figure \ref{fig:collimation} shows the collimation of the shell for parameters $\alpha=0.1$, $\beta=21$, $r_{s0}(0)=10^{-4}$, and different values of the anisotropy parameter $B$ and different values of the exponent $n$ from an isotropic stellar wind ($B=0$), to very anisotropic stellar winds ($B=20$ and $n=4$). As expected, the collimation increases with both $B$ and $n$.

\section{Discussion}
\label{sec:discussion}

We find that at the pole the shell collapses for isotropic and anisotropic stellar winds if the value of the ratio 
between the stellar wind and the accretion flow momentum rates is less than a critical value, 
$\beta < \beta_{\rm crit}$ (see appendix \ref{app:pole}).
This happens because the shell does not have enough momentum to escape. 
At the equator, for a given value of $\beta$, the shell
will   always stagnate near the centrifugal radius  (see appendix \ref{app:equator}). This happens  because the density diverges at $R_{\rm cen}$.


We have also compared the results of our model with those of \citet{Wilkin_2003}. 
 For an isotropic wind and the same values of the parameters $\alpha$ and $\beta$, we find that our shells have the same sizes at the pole. 
Also, the collimation factor  is the same as in their model $C \sim 1.6$ (see appendix \ref{app:wilkin}). The only differences are due to the assumption of 
different  BCs at the disk surface, close to the equator.

The collimation of the shells depends on the anisotropy of the stellar wind and the accretion flow.
In the models with an anisotropic stellar wind and the Ulrich accretion flow,
 it is difficult to get collimation factors $C$ much larger than 3 (figure \ref{fig:collimation}), while the
collimation factors of observed sources have values $\sim$ 3 - 10, \citet{Bontemps_1996}.

We also compare our model with the molecular outflow CB 26 \citep{Launhardt_2009}, at the time
when they both have the same size in the polar direction. The result of this comparison is:
\begin{enumerate}
\item The dynamical time of the model is $t=$ 250 yr which is half of the kinematic age calculated with the observed size
and current velocity. This discrepancy is due to the fact that the shell is decelerating.
\item The radial velocity of the model, is of the order to 10 km s$^{-1}$, consistent with the observed expansion velocity (e.g. \citealt{Lee_2018}).
\item The collimation factor is similar to the observed value.
\item The shell mass in the model is 2$\times$10$^{-3}$ M$_\odot$, twice the observed value.
\item The rotation velocity of the model is lower by an order of magnitude than the observed value (see Table \ref{tab:outflows}).
\item The total angular momentum is also  lower than the observed value.
\end{enumerate}

The low rotation of the model may be resolved, if that the stellar wind has angular momentum, 
or the parent cloud  has more angular momentum than the Ulrich's flow, or with a combination of both mechanisms. 
Part of the problem is that the accreting envelope
does not have large rotation velocities. In addition, the model shell has more mass than the observed shell, 
with slowly rotating material.

Finally, we note that in our thin shell model, we assume that the pressure effects are negligible. Pressure gradients will
not affect the thin shell approximation which depends on an efficient cooling of the shocked gas. 
On the other hand, pressure gradients inside the shell
could change the gas tangential dynamics, accelerating or decelerating the flow along the shell. This effect  is not expected to be important when the flow is supersonic. Thus, to evaluate the effect of the pressure, 
one has  to calculate
the temperature of the shell, which is out of the scope of this paper. 

\section{Conclusions}
\label{sec:conclusions}

To understand the evolution and properties of molecular outflows we 
developed a model of the interaction between a stellar wind and an accretion flow that follows the evolution of a thin shell
that is pushed by and entrains material from both flows.  We have formulated the problem in such a way that we can consider an accretion flow and a stellar wind with general velocity fields (collimated or not collimated and with or without rotation) provided that they have axial symmetry. 

In the present paper we have considered isotropic and anisotropic stellar winds with only radial velocity. The accretion flow is given by the collapse of a slowly rotating molecular cloud from \citet{Ulrich_1976}. The evolution of the outflow was followed from its origin, close to the stellar surface, to large distances from the central star.

The shell evolution has a strong dependence on the ratio between the wind and the accretion flow mass and momentum rates,
$\alpha$ and $\beta$ (eqns. [\ref{eq:alpha}] and [\ref{eq:beta}]). In order for the molecular outflow shells to 
expand, it is necessary that $\beta\ge\beta_{crit}$ for a given value of $\alpha$. If $\beta<\beta_{crit}$, the  whole shell collapses back
to stellar surface.

The interacting flows considered in this work, produce moderate outflow collimation ($C\sim3$) and low rotation ($v_\phi\sim0.1$ km s$^{-1}$). These values are lower than observed in the sources in Table \ref{tab:outflows}. It is left as future work to explore  
other physical collapsing
envelopes and stellar winds to determine the outflow characteristics. 

\medskip
{\it Acknowledgements.} 
J. A. L\'opez-V\'azquez and S. Lizano acknowledge support from PAPIIT-UNAM IN101418 and CONACyT 23863. J. Cant\'o acknowledges support from PAPIIT-UNAM-IG 100218.
We thank an anonymous referee for useful suggestions that improved the presentation of this paper.

\appendix
\section{Derivation of the equations}
\label{app:derivation}
In spherical coordinates the continuity equation is given by
\begin{eqnarray}
\frac{\partial \rho}{\partial t}+\frac{1}{r^2}\frac{\partial(\rho v_r r^2)}{\partial r}+\frac{1}{r \sin\theta}\frac{\partial(\rho v_\theta\sin\theta)}{\partial \theta}+\frac{1}{r\sin\theta}\frac{\partial (\rho v_\phi)}{\partial\phi}=0,
\label{eq:continuity}
\end{eqnarray}

\noindent where $\rho$ is the mass volume density and $v_r$, $v_\theta$, and $v_\phi$ are the fluid velocities. We assume axisimmetry with respect to $\phi$ direction and multiply the above equation by $r^2\sin\theta$. Then, the continuity equation can be written as
\begin{eqnarray}
\frac{\partial(\rho r^2\sin\theta)}{\partial t}+\frac{\partial(\rho  v_r r^2 \sin\theta)}{\partial r}+ \frac{\partial(\rho v_\theta r\sin\theta)}{\partial \theta}=0.
\label{eq:continuity1}
\end{eqnarray}

\noindent Integrating the latter equation in the radial direction from $R$ to $R+\Delta R$ for fixed $\theta$ (see Figure \ref{fig:integration_schemes}), the continuity equation of the shell can be written as

\begin{eqnarray}
\frac{\partial}{\partial t}\left(R_s^2 \sin\theta \sigma\right)+\frac{\partial}{\partial \theta}\left( R_s\sin\theta \sigma U_{\theta}\right)+R_s^2\sin\theta\left[ \rho_a (U_{ar}-U_{r})-\rho_w(U_{wr}-U_{r})\right]=0.
\label{eq:appmass}
\end{eqnarray}

\noindent In this equation the mass surface density of the shell in the radial direction is $\sigma=\rho_s\delta$, and $U_r$ and $U_\theta$ are the radial and $\theta$ velocity components of  the shell material, respectively.

The equation of the fluid momentum in the radial direction is given by
\begin{eqnarray}
\rho\frac{\partial v_r}{\partial t}+\rho v_r\frac{\partial v_r}{\partial r}+\frac{\rho v_\theta}{r}\frac{\partial v_r}{\partial \theta}+\frac{\rho v_\phi}{r\sin\theta}\frac{\partial v_r}{\partial \phi}-\rho\frac{v_\theta^2+v_\phi^2}{r}=F_g,
\label{eq:momentumr}
\end{eqnarray}

\noindent where $F_g=-G M_* \rho/r^2$ is the gravitational force per unit volume.

The equations of the fluid momentum in the $\theta$ and the azimuthal directions are given by 
\begin{eqnarray}
\rho\frac{\partial v_\theta}{\partial t}+\rho v_r\frac{\partial v_\theta}{\partial r}+\frac{\rho v_\theta}{r}\frac{\partial v_\theta}{\partial\theta}+\frac{ \rho v_\phi}{r \sin\theta}\frac{\partial v_\theta}{\partial  \phi}+\rho \frac{v_rv_\theta}{r} -\rho \frac{v_\phi^2\cot\theta}{r}=0,
\label{eq:momentumtheta}
\end{eqnarray}

\noindent and
\begin{eqnarray}
\rho\frac{\partial v_\phi}{\partial t}+\rho v_r\frac{\partial v_\phi}{\partial r}+\frac{\rho v_\theta}{r}\frac{\partial v_\phi}{\partial \theta}+\frac{ \rho v_\phi}{r \sin\theta}\frac{\partial v_\phi}{\partial \phi}+\rho\frac{v_rv_\phi}{r}+\rho\frac{v_\theta v_\phi \cot\theta}{r}=0.
\label{eq:momentumphi}
\end{eqnarray} 

One multiplies eq. (\ref{eq:continuity}) by $v_r$, $v_\theta$, and $v_\phi$, and add the result to eqns. (\ref{eq:momentumr}) - (\ref{eq:momentumphi}), respectively. Then, one integrates each equation in the radial direction for a thin shell, considering the axial symmetry, and obtains the following equations for the momenta of the shell in the radial, the $\theta$ and the azimuthal directions

\begin{eqnarray}
\frac{\partial}{\partial t}\left(R_s^2\sin\theta \sigma U_{r}\right)&+&\frac{\partial}{\partial\theta}\left( R_s\sin\theta \sigma U_{r}U_{\theta}\right)+R_s^2\sin\theta\left[\rho_a U_{ar}(U_{ar}-U_{r})-\rho_w U_{wr}\left(U_{wr}-U_{r} \right)\right]\nonumber \\
&-&R_s\sin\theta \sigma\left(U_{\theta}^2+U_{\phi}^2\right)+GM_*\sin\theta \sigma=0,
\label{eq:appmomr}
\end{eqnarray}

\begin{eqnarray}
\frac{\partial}{\partial t}\left(R_s^2\sin\theta  \sigma U_{\theta}\right)&+&\frac{\partial}{\partial\theta}\left(R_s\sin\theta \sigma U_{\theta} ^2\right) +R_s^2\sin\theta\left[\rho_a U_{a\theta}\left( U_{ar}-U_{r}\right)-\rho_w U_{w\theta}\left(U_{wr}-U_r\right)\right]\nonumber \\
&+&R_s\sin\theta\sigma \left(U_{r}U_{\theta}-U_{\phi}^2\cot\theta\right)=0,
\label{eq:appmomtheta}
\end{eqnarray}

\noindent and
\begin{eqnarray}
\frac{\partial}{\partial t}\left(R_s^2\sin\theta \sigma U_{\phi} \right)&+&\frac{\partial}{\partial\theta}\left(R_s\sin\theta \sigma U_{\theta} U_{\phi} \right)+R_s^2\sin\theta \left[\rho_aU_{a\phi}  \left( U_{ar}-U_{r}\right)-\rho_wU_{w\phi}  \left( U_{wr}-U_{r}\right)\right]\nonumber \\
&+& R_s\sin\theta\sigma\left( U_{r}U_{\phi}+U_{\theta} U_{\phi} \cot\theta\right)=0,
\label{eq:appmomphi}
\end{eqnarray}
where $U_\phi$ is the azimuthal velocity of the shell material.

To obtain the evolution of the shell radius, one can write 
\begin{eqnarray}
 U_r =  {d R_s \over d t} &=&{\partial R_s \over \partial t} + {\partial R_s \over \partial \theta }{\partial \theta \over \partial t} \nonumber \\
&=& {\partial R_s \over \partial t} +  {1 \over R_s} {\partial R_s \over \partial \theta } R_s {\partial \theta \over \partial t} \nonumber \\
&=& {\partial R_s \over \partial t} +  {U_\theta \over R_s} {\partial R_s \over \partial \theta } .
\end{eqnarray}
Then, the evolution of the radius of the shell is given by
\begin{eqnarray}
\frac{\partial R_s}{\partial t}=U_{r}-\frac{U_\theta}{R_s}\frac{\partial R_s}{\partial \theta}.
\label{eq:rs}
\end{eqnarray}

Eqns. (\ref{eq:appmass}), (\ref{eq:appmomr}), (\ref{eq:appmomtheta}), (\ref{eq:appmomphi}), and (\ref{eq:rs}) describe the evolution of the shell formed by the shock between a stellar wind and an accretion flow with a given velocity field. These equations can be written the more compact form shown in the eqns. (\ref{eq:pmass}) - (\ref{eq:prshell}) in terms of the mass and momentum fluxes (eqns. [\ref{eq:pm}] and [\ref{eq:prtp}]).

\section{Radius and velocities around the pole}
\label{app:pole}

Around the pole, the mass and the momentum fluxes, and the radius can be expanded as power series of $\theta$ to second order as
\begin{eqnarray}
p_m\approx b_{m1}\theta+b_{m2}\theta^2,
\label{eq:apppolepm}
\end{eqnarray}

\begin{eqnarray}
p_r\approx b_{r1}\theta+b_{r2}\theta^2,
\label{eq:apppolepr}
\end{eqnarray}

\begin{eqnarray}
p_\theta\approx b_{\theta1}\theta+b_{\theta2}\theta^2,
\label{eq:apppoleptheta}
\end{eqnarray}

\begin{eqnarray}
p_\phi\approx b_{\phi1}\theta+b_{\phi2}\theta^2,
\label{eq:apppolepphi}
\end{eqnarray}

\noindent and
\begin{eqnarray}
r_s\approx r_{s0}+r_{s1}\theta+r_{s2}\theta^2.
\label{eq:apppolers}
\end{eqnarray}

\noindent In the above equations, we can note that the mass and the momentum fluxes do not have component of order zero, since these momenta are zero at $\theta=0$.

Substituting the eq. (\ref{eq:polers}) in eq. (\ref{eq:zeta}) and expanding in Taylor's series for $\theta\ll1$ and $\theta_0\ll1$, one finds the relation between the radius at the pole $r_{s0}$, the angle $\theta$, and the angle $\theta_0$,
\begin{eqnarray}
\theta_0\approx\left(\frac{r_{s0}}{2+r_{s0}}\right)^{1/2}\theta.
\label{eq:theta0pole}
\end{eqnarray}

In addition, the velocities and the density of the accretion flow are expanded in Taylor's series to first order in $\theta$. 
Using the latter equation for $\theta_0$ one obtains

\begin{eqnarray}
u_{ar}\approx-\left(\frac{2}{r_{s0}}\right)^{1/2},
\label{eq:apppoleuar}
\end{eqnarray}

\begin{eqnarray}
u_{a\theta}\approx\left(\frac{2}{r_{s0}}\right)^{1/2}\frac{1}{2+r_{s0}}\theta,
\label{eq:apppoleuat}
\end{eqnarray}

\begin{eqnarray}
u_{a\phi}\approx-\frac{1}{2+r_{s0}}\theta,
\label{eq:apppoleuap}
\end{eqnarray}

\begin{eqnarray}
\rho^\prime_a\approx\left(\frac{1}{2r_{s0}}\right)^{1/2}\frac{1}{2+r_{s0}}.
\label{eq:apppolerhoa}
\end{eqnarray}
Also, the expansion for the wind density gives
\begin{eqnarray}
\rho^\prime_w \approx \left[\frac{A+B}{A+B/(2n+1)}\right]\left(\frac{\alpha}{r_{s0}^2 u_w}\right).
\label{eq:apprhowp}
\end{eqnarray}

Substituting the eqns. (\ref{eq:apppolepm}) - (\ref{eq:apprhowp}) in eqns. (\ref{eq:tildepm}) - (\ref{eq:trs}), one finds that the coefficients $b_{m2}(\tau)=b_{r2}(\tau)=b_{\theta1}(\tau)=b_{\phi1}(\tau)=r_{s1}(\tau)=r_{s2}(\tau)=0$. In addition, one obtains a set of ordinary differential equations for the functions $b_{m1}(\tau)$, $b_{r1}(\tau)$, $b_{\theta2}(\tau)$, $b_{\phi2}(\tau)$, and $r_{s0}(\tau)$. These equations are,
\begin{eqnarray}
\frac{d b_{m1}}{d\tau}+\frac{2b_{\theta2}}{r_{s0}}=\alpha\left[\frac{A+B}{A+B/(2n+1)}\right]\left(1-\frac{\alpha}{\beta}\frac{b_{r1}}{b_{m1}}\right)+\left(\frac{r_{s0}}{2+r_{s0}}\right)\left(\frac{r_{s0}}{2}\right)^{1/2}\left[\frac{b_{r1}}{b_{m1}}+\left(\frac{2}{r_{s0}}\right)^{1/2}\right],
\label{eq:polebm}
\end{eqnarray}

\begin{eqnarray}
\frac{db_{r1}}{d\tau}+\frac{2b_{r1}b_{\theta2}}{b_{m1}r_{s0}}+\frac{b_{m1}}{r_{s0}^2}=\beta\left[\frac{A+B}{A+B/(2n+1)}\right]\left(1-\frac{\alpha}{\beta}\frac{b_{r1}}{b_{m1}}\right)-\left(\frac{r_{s0}}{2+r_{s0}}\right)\left[\frac{b_{r1}}{b_{m1}}+\left(\frac{2}{r_{s0}}\right)^{1/2}\right],
\label{eq:polebr}
\end{eqnarray}

\begin{eqnarray}
\frac{db_{\theta2}}{d\tau}-\frac{b_{\phi2}^2}{b_{m1}r_{s0}}+\frac{b_{r1}b_{\theta2}}{b_{m1}r_{s0}}+\frac{3b_{\theta2}^2}{b_{m1}r_{s0}}-\frac{(2r_{s0})^{1/2}}{(2+r_{s0})^2}=\left( \frac{b_{r1}}{b_{m1}}\right)\left[\frac{r_{s0}}{(2+r_{s0})^2}\right],
\label{eq:polebtheta}
\end{eqnarray}

\begin{eqnarray}
\frac{db_{\phi2}}{d\tau}+\frac{b_{\phi2}b_{r1}}{b_{m1}r_{s0}}+\frac{4b_{\theta2}b_{\phi2}}{b_{m1}r_{s0}}+\frac{r_{s0}}{(2+r_{s0})^2}=-\left(\frac{b_{r1}}{b_{m1}}\right)\left(\frac{r_{s0}}{2}\right)^{1/2}\left[\frac{r_{s0}}{(2+r_{s0})^2}\right],
\label{eq:polebphi}
\end{eqnarray}

\noindent and
\begin{eqnarray}
\frac{d r_{s0}}{d\tau}=\frac{b_{r1}}{b_{m1}}.
\label{eq:dbrs0}
\end{eqnarray}

\noindent The coefficients $b_{m1}(\tau)$, $b_{r1}(\tau)$, $b_{\theta2}(\tau)$, $b_{\phi2}(\tau)$, and $r_{s0}(\tau)$ are functions of the time ($\tau$). The shell starts at an initial radius $r_{s0}(0)$, close to the stellar radius.
Because initially the shell is massless,  $b_{m1}(0)=b_{r1}(0)=b_{\theta2}(0)=b_{\phi2}(0)=0$. Thus, to find the ratios of $b_{r1}/b_{m1}$, $b_{\theta2}/b_{m1}$, and $b_{\phi2}/b_{m1}$, we expand the coefficients to first order in $\tau$ for $\tau\ll1$,
\begin{eqnarray}
b_{m1}\approx c_m\tau,
\label{eq:bmt0}
\end{eqnarray}

\begin{eqnarray}
b_{r1}\approx c_r\tau,
\label{eq:brt0}
\end{eqnarray}

\begin{eqnarray}
b_{\theta2}\approx c_\theta \tau,
\label{eq:btt0}
\end{eqnarray}

\begin{eqnarray}
b_{\phi2}\approx c_\phi \tau,
\label{eq:bpt0}
\end{eqnarray}

\noindent and
\begin{eqnarray}
r_{s0}\approx r_{s0}(0)+c_{rs}\tau.
\label{eq:rt0}
\end{eqnarray}

Substituting these equations in eqns. (\ref{eq:polebm}) - (\ref{eq:dbrs0}), one obtains $c_m$, $c_r$, $c_\theta$, $c_\phi$, and $c_{rs}$ as  function of the initial radius $r_{s0}(0)$ and the ratio 
\begin{eqnarray}
\lambda=\frac{c_r}{c_m}.
\label{eq:lambda}
\end{eqnarray}
Then, 
\begin{eqnarray}
c_m=-\alpha\left[\frac{A+B}{A+B/(2n+1)}\right]\left(\frac{\alpha}{\beta}\lambda-1\right)+Q_1\left(\frac{r_{s0}(0)}{2}\right)^{1/2}\left[\lambda+\left(\frac{2}{r_{s0}(0)}\right)^{1/2}\right],
\label{eq:cm}
\end{eqnarray}

\begin{eqnarray}
c_r=-\beta\left[\frac{A+B}{A+B/(2n+1)}\right]\left(\frac{\alpha}{\beta}\lambda-1\right)+Q_1\left[\lambda+\left(\frac{2}{r_{s0}(0)}\right)^{1/2}\right],
\label{eq:cr}
 \end{eqnarray}

\begin{eqnarray}
c_\theta=\frac{Q_1}{2+r_{s0}(0)}\left[\lambda+\left(\frac{2}{r_{s0}(0)}\right)^{1/2}\right],
\label{eq:ct}
\end{eqnarray}

\begin{eqnarray}
c_\phi=-\frac{Q_1}{2+r_{s0}(0)}\left(\frac{r_{s0}(0)}{2}\right)^{1/2}\left[\lambda+\left(\frac{2}{r_{s0}(0)}\right)^{1/2}\right],
\label{eq:cp}
\end{eqnarray}

\begin{eqnarray}
c_{rs}=\lambda,
\label{eq:cr1}
\end{eqnarray}

\noindent where $Q_1$ is given by,

\begin{eqnarray}
Q_1=\frac{r_{s0}(0)}{2+r_{s0}(0)}.
\label{eq:q1}
\end{eqnarray}

\noindent One also obtains a quadratic equation for $\lambda$ substituting eqns. (\ref{eq:cm}) and (\ref{eq:cr}) in  
eq. (\ref{eq:lambda}),
\begin{eqnarray}
&&\left[\frac{\alpha^2}{\beta}\left(\frac{A+B}{A+B/(2n+1)}\right)-\left(\frac{r_{s0}(0)}{2}\right)^{1/2}Q_1\right]\lambda^2-2\left[\alpha\left(\frac{A+B}{A+B/(2n+1)}\right)+Q_1\right]\lambda \nonumber \\
&+&\beta\left(\frac{A+B}{A+B/(2n+1)}\right)-\left(\frac{2}{r_{s0}(0)}\right)^{1/2}Q_1=0.
\label{eq:lambda2}
\end{eqnarray}

Eqns. (\ref{eq:polebm}) - (\ref{eq:dbrs0}) are solved numerically 
for the coefficients  $b_{m1}$, $b_{r1}$, $b_{\theta2}$, $b_{\phi2}$, and $r_{s0}$. The 
initial conditions are obtained from eqns. (\ref{eq:bmt0}) - (\ref{eq:rt0}). The latter are evaluated at a very small non dimensional time,  $\tau=10^{-9}$.  From the solution of these equations one finds BCs at the pole for the mass and momentum fluxes, and the shell radius as a function of time.

\begin{figure}[h!]
\centering
\includegraphics[scale=0.5]{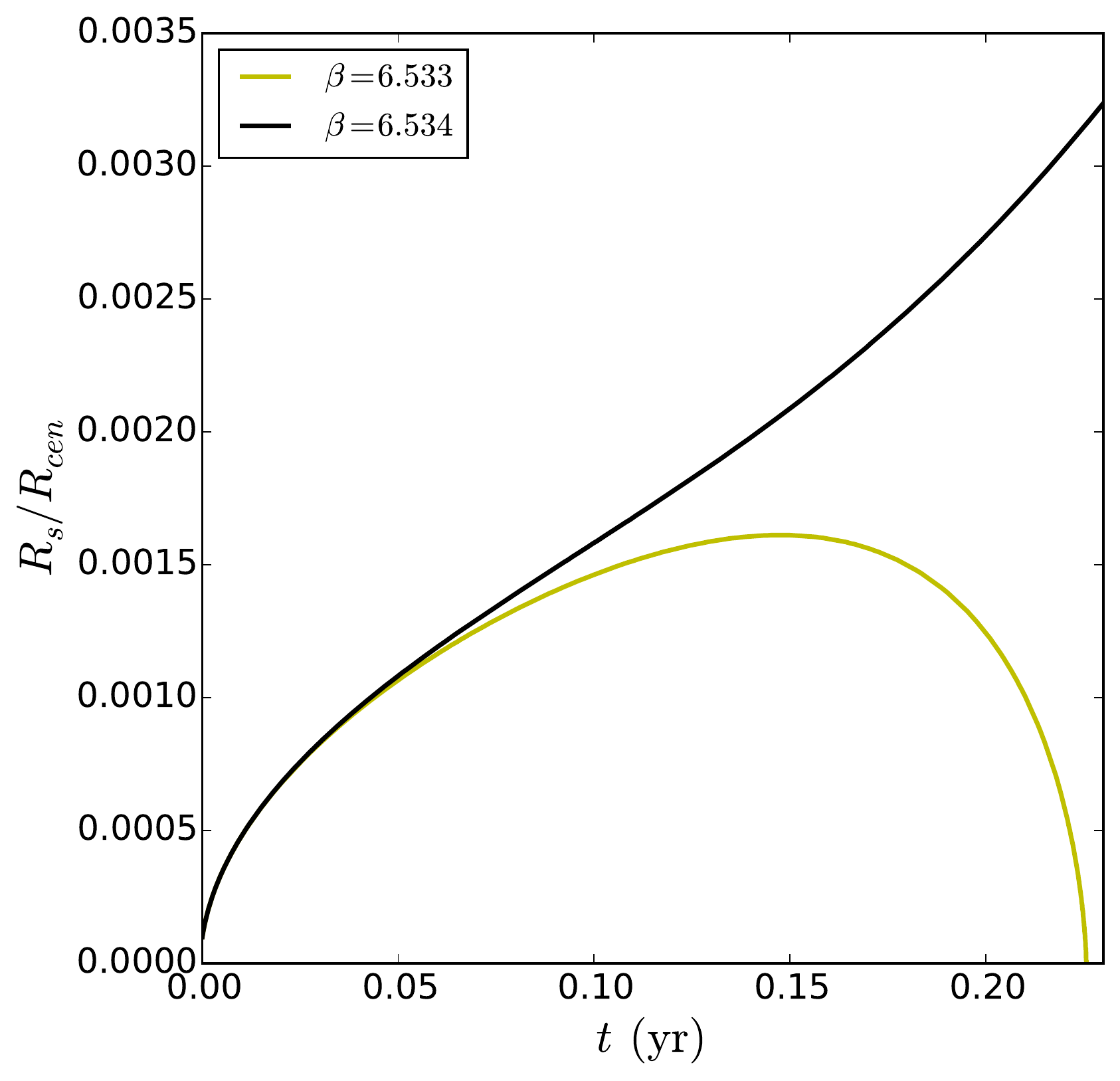}
\caption{Evolution of the shell radius $R_s$ at the pole for an isotropic stellar wind ($B=0$) and parameters $\alpha=0.1$ and an initial radius $r_{s0}(0)=10^{-4}$. The shell expands for $\beta_{crit}=6.534$ (black line); and the shell collapses for $\beta<\beta_{crit}$, yellow line (see Table \ref{tab:bcritpole}).}
\label{fig:rpolo}
\end{figure}

For the case of the isotropic stellar wind ($B=0$ in eq. [\ref{eq:fanisotropic}]), we explore the critical value of $\beta$ required for the expansion of the shell. Table \ref{tab:bcritpole} shows the critical value $\beta_{crit}$, for different values of $\alpha$ and initial radius $r_{s0}(0)$. 
Figure \ref{fig:rpolo} shows the shell radius at the pole for early times for a model with 
$\alpha=0.1$, initial radius $r_{s0}(0)=10^{-4}$, and two cases:  $\beta=\beta_{crit}$ and $\beta<\beta_{crit}$. In the former case, the shell always expands; in the latter case the shell collapses back onto the stellar surface.

To obtain $\beta_{crit}$, one takes into account the weight of the shell, the change of the radial momentum in the $\theta$ direction, and the momentum added to the shell by  both the stellar wind and the accretion flow. If one neglects the weight of the shell for the models in Table \ref{tab:bcritpole}, one obtains that a smaller value $\beta > 0.5$ is enough for the shell to expand.

\begin{figure*}[h!]
\centering
\includegraphics[scale=0.5]{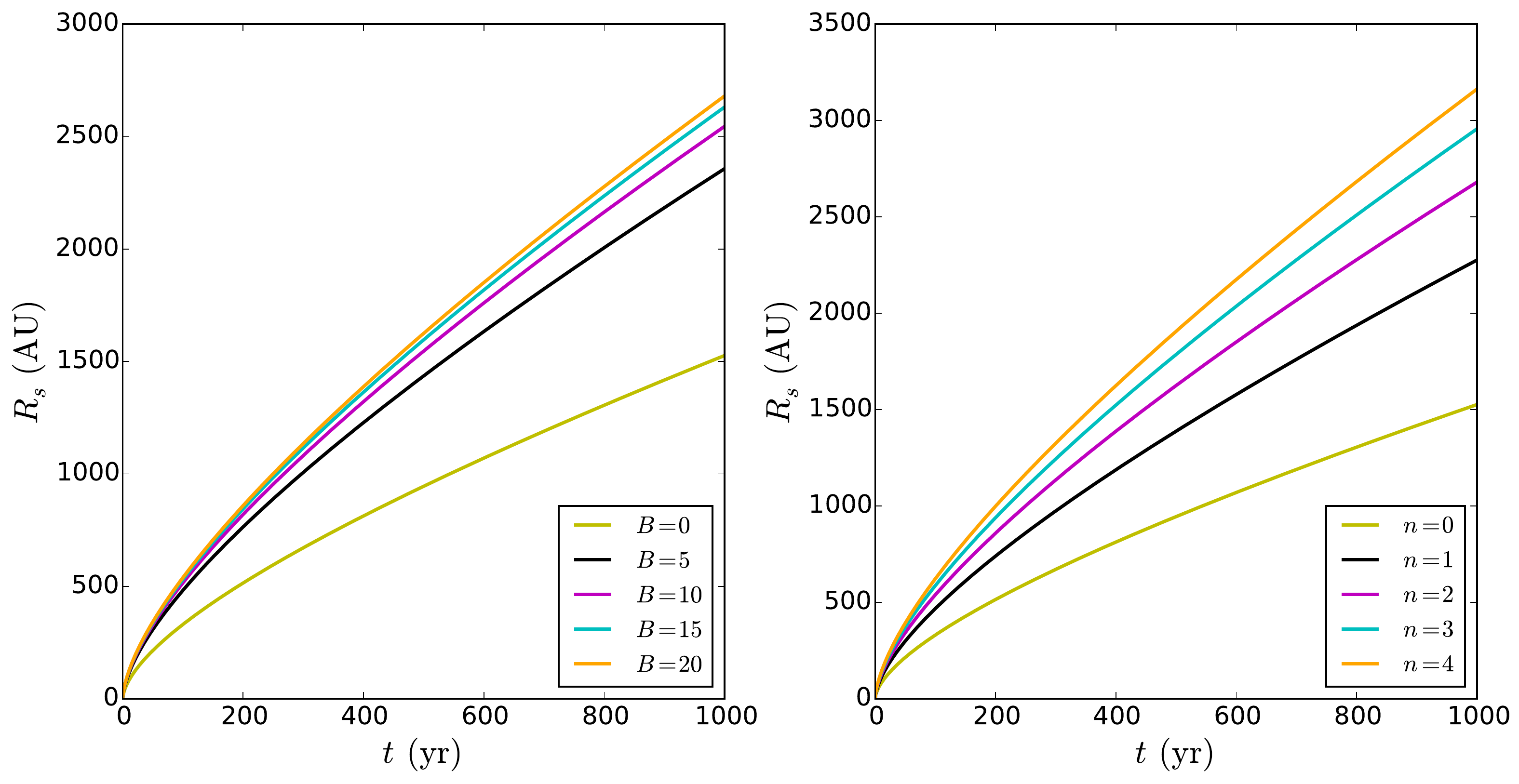}
\caption{Radius of the shell at the pole $\theta=0$ as a function of time for the parameters $\alpha=0.1$, $\beta=21$, and $r_{s0}(0)=10^{-4}$. Left panel: stellar winds with $A=1$, $n=2$, and different values of the anisotropy parameter $B=0$, 5, 10, 15, and 20. Right panel: anisotropic stellar winds with $A=1$, $B=20$, and different exponents $n=0$, 1, 2, 3, and 4.}
\label{fig:rpolea}
\end{figure*}

\begin{figure}[h!]
\centering
\includegraphics[scale=0.5]{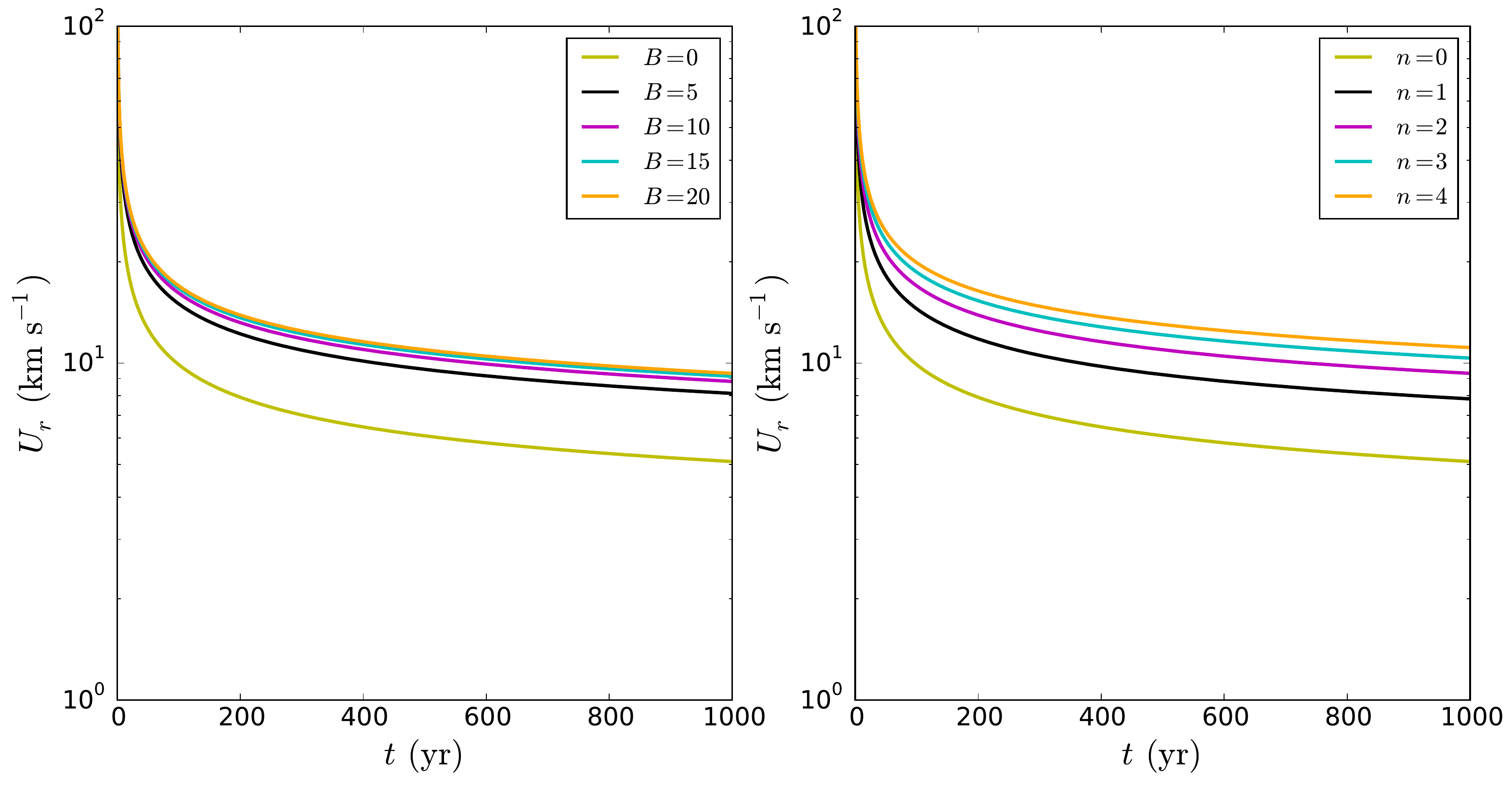}
\caption{Radial velocity of the shell at the pole $\theta=0$ as a function of time for the parameters $\alpha=0.1$, $\beta=21$, and $r_{s0}(0)=10^{-4}$. Left panel: stellar winds with $A=1$, $n=2$, and different values of the anisotropy parameter $B=0$, 5, 10, 15, and 20. Right panel: anisotropic stellar winds with $A=1$, $B=20$, and different exponents $n=0$, 1, 2, 3, and 4.}
\label{fig:urpolo}
\end{figure}

Assuming $\alpha=0.1$ and $\beta=21$, 
Figure \ref{fig:rpolea} shows the evolution of the shell radius for models with parameters $A=1$ and $n=2$, and 
 different values of the anisotropy parameter $B$  (left panel), and models with parameters $A=1$ and $B=20$, 
 and different values of the exponent $n$ (right panel). This figure shows that  the shell radius increases with the anisotropy parameter $B$ and the exponent $n$. For the same models, 
Figure \ref{fig:urpolo} shows the radial velocity at the pole (the shell expansion velocity).
The radial velocity also increases with the anisotropy parameter $B$ and the exponent $n$, and it tends to a constant value at large times. 
The $\theta$ and the azimuthal velocities at the pole are zero because 
$v_\theta = (b_{\theta 2}/ b_{m1}) \theta$ and $v_\phi = (b_{\phi 2}/ b_{m1}) \theta$ are linear functions of $\theta$.

\begin{table}[h!]
\centering
\caption{Values of $\beta_\mathrm{crit}$ for different values of $\alpha$ and initial radius $r_{s0}(0)$.}
\begin{tabular}{ c | c c c}
\hline
\hline
\diagbox[innerwidth=1cm,innerleftsep=-0.75cm,innerrightsep=-5pt]{$\alpha$}{$r_{s0}(0)$} & $10^{-5}$ &5$\times10^{-5}$& $10^{-4}$ \\
\hline
0.01 & 2.061 & 0.972 & 0.900 \\
0.10 & 20.084 & 9.140 & 6.534 \\
0.50 & 99.732 & 45.085 &  32.023 \\
\hline
\end{tabular}
  \label{tab:bcritpole}  
\end{table}

\section{Radius and velocities around the equator}
\label{app:equator}

Around the equator the mass and the momentum fluxes, and the radius are expanded as power series around the angle $\left(\frac{\pi}{2}-\eta\right)-\theta$. These expansions are
\begin{eqnarray}
p_m\approx q_{m0}+q_{m1}\left[\left(\frac{\pi}{2}-\eta\right)-\theta\right],
\label{eq:apppmeq}
\end{eqnarray}

\begin{eqnarray}
p_r\approx q_{r0}+q_{r1}\left[\left(\frac{\pi}{2}-\eta\right)-\theta\right],
\label{eq:apppreq}
\end{eqnarray}

\begin{eqnarray}
p_\theta\approx q_{\theta0}+q_{\theta1}\left[\left(\frac{\pi}{2}-\eta\right)-\theta\right],
\label{eq:apppteq}
\end{eqnarray}

\begin{eqnarray}
p_\phi\approx q_{\phi0}+q_{\phi1}\left[\left(\frac{\pi}{2}-\eta\right)-\theta\right],
\label{eq:appppeq}
\end{eqnarray}

\noindent and
\begin{eqnarray}
r_s\approx q_{rs0}+q_{rs1}\left[\left(\frac{\pi}{2}-\eta\right)-\theta\right],
\label{eq:appradeq}
\end{eqnarray}
where $\left(\frac{\pi}{2}-\eta\right)-\theta\ll1$.

The expansion in Taylor's series to first order around the angle $\left(\frac{\pi}{2}-\eta\right)-\theta$ of the velocity field and the density of the accretion flow
 gives
\begin{eqnarray}
u_{ar}\approx-\left[\left(\frac{1}{q_{rs0}}\right)\left(1+\frac{\sin\eta}{\cos\theta_0}\right)\right]^{1/2},
\label{eq:appuareq}
\end{eqnarray}
\begin{eqnarray}
u_{a\theta}\approx\left[\left(\frac{1}{q_{rs0}}\right)\left(1+\frac{\sin\eta}{\cos\theta_0}\right)\right]^{1/2}\left(\frac{\cos\theta_0-\sin\eta}{\cos\eta}\right),
\label{eq:appuateq}
\end{eqnarray}
\begin{eqnarray}
u_{a\phi}\approx-\left[\left(\frac{1}{q_{rs0}}\right)\left(1-\frac{\sin\eta}{\cos\theta_0}\right)\right]^{1/2}\left(\frac{\sin\theta_0}{\cos\eta}\right),
\label{eq:appuapeq}
\end{eqnarray}
and
\begin{eqnarray}
\rho^\prime_{a}\approx\frac{1}{(q_{rs0}-1+3\cos^2\theta_0)\left(1+\frac{\sin\eta}{\cos\theta_0}\right)^{1/2}}\left(\frac{1}{q_{rs0}}\right)^{1/2},
\label{eq:apprhoaeq}
\end{eqnarray}
where $\theta_0$ is obtained of the eq. (\ref{eq:zeta}).

The wind density is
\begin{eqnarray}
\rho^\prime_{w}\approx\left[\frac{A+B\sin^{2n}\eta}{A+B/(2n+1)}\right]\frac{\alpha^2}{q_{rs0}^2 u_{wr}}.
\label{eq:apprhoweq}
\end{eqnarray}

Substituting eqns. (\ref{eq:apppmeq}) - (\ref{eq:appradeq}), the velocity field and the density of the accretion flow, and the density of the stellar wind given in eqns. (\ref{eq:appuareq}) - (\ref{eq:apprhoweq}), in eqns. (\ref{eq:tildepm}) - (\ref{eq:trs}), one obtains a set of differential equations 
for the coefficients
\begin{eqnarray}
\frac{dq_{m0}}{d\tau}+f_{m0,1}=f_{m0,2},
\label{eq:dqm0}
\end{eqnarray}

\begin{eqnarray}
\frac{dq_{m1}}{d\tau}+f_{m1,1}=f_{m1,2}.
\label{eq:dqm1}
\end{eqnarray}

\begin{eqnarray}
\frac{dq_{r0}}{d\tau}+f_{r0,1}=f_{r0,2},
\label{eq:dqr0}
\end{eqnarray}

\begin{eqnarray}
\frac{dq_{r1}}{d\tau}+f_{r1,1}=f_{r1,2},
\label{eq:dqr1}
\end{eqnarray}

\begin{eqnarray}
\frac{dq_{\theta0}}{d\tau}+f_{\theta0,1}=f_{\theta0,2},
\label{eq:dqtheta0}
\end{eqnarray}

\begin{eqnarray}
\frac{dq_{\theta1}}{d\tau}+f_{\theta1,1}=f_{\theta1,2},
\label{eq:dqtheta1}
\end{eqnarray}

\begin{eqnarray}
\frac{dq_{\phi0}}{d\tau}+f_{\phi0,1}=f_{\phi0,2},
\label{eq:dqphi0}
\end{eqnarray}

\begin{eqnarray}
\frac{dq_{\phi1}}{d\tau}+f_{\phi1,1}=f_{\phi1,2},
\label{eq:dqphi1}
\end{eqnarray}

\begin{eqnarray}
\frac{dq_{rs0}}{d\tau}=\frac{q_{r0}}{q_{m0}}+\frac{q_{\theta 0}}{q_{m0}}\frac{q_{rs1}}{q_{rs0}},
\label{eq:dqrs0}
\end{eqnarray}
and
\begin{eqnarray}
\frac{dq_{rs1}}{d\tau}=\frac{q_{r0}}{q_{m0}}\left(\frac{q_{r1}}{q_{r0}}-\frac{q_{m1}}{q_{m0}}\right)-\frac{q_{\theta 0}}{q_{m0}}\frac{q_{rs1}}{q_{rs0}}\left(\frac{q_{m1}}{q_{m0}}-\frac{q_{\theta1}}{q_{\theta0}}+\frac{q_{rs1}}{q_{rs0}}\right),
\label{eq:dqrs1}
\end{eqnarray}
where the functions $f_{ai,j}$ are given by
\begin{eqnarray}
f_{m0,1}=P_2 T_3,
\label{eq:fm01}
\end{eqnarray}
\begin{eqnarray}
f_{m0,2}=-\cos\eta\left[\alpha\left(\frac{A+B\sin\eta}{A+B/(2n+1)}\right)\left(\frac{\alpha}{\beta}P_1-1\right)-\frac{q_{rs0}^{3/2}P_3P_4}{\Gamma_2}\right],
\label{eq:fm01}
\end{eqnarray}
\begin{eqnarray}
f_{m1,1}&=&q_{rs0}P_4\cos\eta\left[\frac{q_{rs0}^{1/2}}{\Gamma_2}\left(\frac{T_1\Gamma_2}{2q_{rs0}^{1/2}}+\frac{\sin\eta\Gamma_4}{2q_{rs0}^{1/2}\Gamma_1^2\Gamma_2}-P_1T_4-\frac{\cos\eta}{2q_{rs0}^{1/2}\Gamma_1\Gamma_2}\right)\right. \nonumber \\
&+&\left.\frac{P_3}{2\Gamma_2}\left(-q_{rs0}^{1/2}T_1 -\frac{q_{rs0}^{1/2}}{\Gamma_1\Gamma_2^2}\left(\frac{\sin\eta\Gamma_4}{\Gamma_1}-\cos\eta\right)\right) \right],
\label{eq:fm11}
\end{eqnarray}
\begin{eqnarray}
f_{m1,2}&=&-\cos\eta\left[\frac{\alpha}{A+B/(2n+1)}\left(B\cos\eta\left(\frac{\alpha}{\beta}P1-1\right)+\frac{\alpha}{\beta}P_1T_4\left(B\sin\eta+1\right)\right)+\frac{q_{rs0}^{3/2}P_3P_4^2}{\Gamma2}\left(T_1\left(1-3\Gamma_1^2\right)+6\Gamma_1\Gamma_4\right)\right]\nonumber \\
&+&2P_2T_1T_3+\sin\eta\left[\alpha\left(\frac{A+B\sin\eta}{A+B/(2n+1)}\right)\left(\frac{\alpha}{\beta}P_1-1\right)-\frac{q_{rs0}^{3/2}P_3P_4}{\Gamma2}\right],
\label{eq:fm12}
\end{eqnarray}
\begin{eqnarray}
f_{r0,1}=\frac{q_{m0}}{q_{rs0}^2}-\frac{q_{\theta0}^2+q_{\phi0}^2}{q_{m0}q_{rs0}}+P_1P_2\left(T_3-T_4\right),
\label{eq:fr01}
\end{eqnarray}
\begin{eqnarray}
f_{r0,2}=-\cos\eta\left[\beta\left(\frac{A+B\sin\eta}{A+B/(2n+1)}\right)\left(\frac{\alpha}{\beta}P_1-1\right)+q_{rs0}P_3P_4\right],
\label{eq:fr02}
\end{eqnarray}
\begin{eqnarray}
f_{r1,1}&=&\frac{q_{m0}}{q_{rs0}^2}\left(T_2-3T_1\right)+q_{r0}T_8+2P_1P_2T_3T_5+\frac{T_2\left(q_{\theta0}^2+q_{\phi0}^2\right)-2\left(q_{\theta0}q_{\theta1}+q_{\phi0}q_{\phi1}+q_{r1}q_{\theta1}\right)}{q_{m0}q_{rs0}},
\label{eq:fr11}
\end{eqnarray}
\begin{eqnarray}
f_{r1,2}&=&\cos\eta\left[-\frac{\beta}{1+A/(2n+1)}\left(A\cos\eta\left(\frac{\alpha}{\beta}P_1-1\right)+\frac{\alpha}{\beta}P_1T_4\left(1+A\sin\eta\right)\right)+q_{rs0}P_3P_4^2\left(T_1(1-3\Gamma_1^2)+6\Gamma_1\Gamma_4\right)\right]\nonumber \\
&-&\frac{q_{rs0}^{1/2}P_4\cos\eta}{2}\left(T_1\Gamma_2+\frac{\sin\eta\Gamma_4}{\Gamma_1^2\Gamma_2}-P_1T_4-\frac{\cos\eta}{\Gamma_1\Gamma_2}\right)+\sin\eta\left[\beta\left(\frac{1+A\sin\eta}{1+A/(2n+1)}\right)\left(\frac{\alpha}{\beta}P_1-1\right)+q_{rs0}P_3P_4\right],
\label{eq:fr12}
\end{eqnarray}
\begin{eqnarray}
f_{\theta0,1}=P_2\left(P_1+\frac{q_{\theta0}}{q_{m0}}T_2\right)-\frac{2q_{\theta0}q_{\theta1}+q_{\phi0}^2\tan\eta}{q_{m0}q_{rs0}},
\label{eq:ftet01}
\end{eqnarray}
\begin{eqnarray}
f_{\theta0,2}=\left(\Gamma_1-\sin\eta\right)q_{rs0}P_3P_4,
\label{eq:ftet02}
\end{eqnarray}
\begin{eqnarray}
f_{\theta1,1}&=&\frac{4q_{\theta0}q_{\theta1}T_2-\left(2q_{\theta1}^2+q_{\phi0}^2\right)+\tan\eta\left[q_{\phi0}^2\left(T_2-\tan\eta\right)-2q_{\phi0}q_{\phi1}\right]}{q_{m0}q_{rs0}}\nonumber \\
&+&q_{rs0}P_3P_4^2\left(\Gamma_1-\sin\eta\right)\left[T_1\left(1-3\Gamma_1^2\right)+6\Gamma_1\Gamma_4\right],
\label{eq:ftheta11}
\end{eqnarray}
\begin{eqnarray}
f_{\theta1,2}&=&P_1P_2\left(T_2-T_5\right)-q_{\theta0}T_8-q_{rs0}P_4\left[P_3\left(\cos\eta-\Gamma_4\right)+\frac{\Gamma_1-\sin\eta}{2q_{rs0}^{1/2}}\left(T_1\Gamma_2+\frac{\sin\eta\Gamma4}{\Gamma_1^2\Gamma_2}-P_1T_4-\frac{\cos\eta}{\Gamma_1\Gamma_2}\right)\right],
\label{eq:ftheta12}
\end{eqnarray}
\begin{eqnarray}
f_{\phi0,1}=q_{\phi0}\left[\frac{P_1}{q_{rs0}}+\frac{P_2}{q_{m0}}\left(T_2-T_7+\tan\eta\right)\right],
\label{eq:fphi01}
\end{eqnarray}
\begin{eqnarray}
f_{\phi0,2}=-\frac{q_{rs0}P_3P_4 \Gamma_3\sin\theta_0}{\Gamma_2},
\label{eq:fphi02}
\end{eqnarray}
\begin{eqnarray}
f_{\phi1,1}=q_{\phi0}\left(T_8+\frac{2P_2T_2T_7}{q_{m0}}\right)+\frac{q_{\theta0}q_{\phi0}-q_{r0}q_{\phi0}\left(T_2-T_6\right)-2q_{\theta1}q_{\phi1}}{q_{m0}q_{rs0}}+\frac{q_{\phi0}P_2\tan\eta}{q_{m0}}\left(T_7-T_2+\tan\eta\right),
\label{eq:fphi11}
\end{eqnarray}
and
\begin{eqnarray}
f_{\phi1,2}&=&\frac{q_{rs0}\sin\theta_0P_3P_4^2}{\Gamma_2}\left[T_1\left(1-3\Gamma_1^2\right)+6\Gamma_1\Gamma_4\right]+\frac{q_{rs0}\sin\theta_0P_4\Gamma_3}{2q_{rs0}^{1/2}\Gamma_2}\left(T_1\Gamma_2+\frac{\sin\eta\Gamma_4}{\Gamma_1^2\Gamma_2}-2q_{rs0}^{1/2}P_1T_4-\frac{\cos\eta}{\Gamma_1\Gamma_2}\right)\nonumber \\
&+&q_{rs0}P_3P_4\left[\frac{\Gamma_1\Gamma_3\Gamma_4}{\sin\theta_0\Gamma_2}-\frac{\sin\theta_0}{2\Gamma_1\Gamma_2}\left(\frac{\Gamma_3}{\Gamma_2^2}+\frac{1}{\Gamma_3}\right)\left(\frac{\sin\eta\Gamma_4}{\Gamma_1}-\cos\eta\right)\right].
\label{eq:fphi12}
\end{eqnarray}

In these equations the functions $\Gamma_i$ are defined as,
\begin{eqnarray}
\Gamma_1=\cos\theta_0,
\label{eq:g1}
\end{eqnarray}

\begin{eqnarray}
\Gamma_2=\left(1+\frac{\sin\eta}{\cos\theta_0}\right)^{1/2},
\label{eq:g2}
\end{eqnarray}

\begin{eqnarray}
\Gamma_3=\left(1-\frac{\sin\eta}{\cos\theta_0}\right)^{1/2},
\label{eq:g3}
\end{eqnarray}
and
\begin{eqnarray}
\Gamma_4=q_{rs1}\frac{\partial \cos\theta_0}{\partial r}-\frac{\partial \cos\theta_0}{\partial \theta}.
\label{eq:g4}
\end{eqnarray}
In the above equation, the partial derivatives of $\cos\theta_0$ are obtained writing eq. (\ref{eq:zeta})
as 
\begin{eqnarray}
\cos^3\theta_0+\left(\frac{1}{\zeta}-1\right)\cos\theta_0-\frac{1}{\zeta}\cos\theta=0,
\label{eq:soltheta0}
\end{eqnarray}
where $\zeta=1/r_{s}$, and are given by

\begin{eqnarray}
\frac{\partial \cos\theta_0}{\partial r_s}=\frac{\cos\theta-\cos\theta_0}{\cos^2\theta_0+r_s-1},
\end{eqnarray}

\begin{eqnarray}
\frac{\partial \cos\theta_0}{\partial \theta}=-\frac{r_s\sin\theta}{3\cos^2\theta_0+r_s-1}.
\end{eqnarray}

The functions $P_i$ are given by
\begin{eqnarray}
P_1=\frac{q_{r0}}{q_{m0}},
\end{eqnarray}

\begin{eqnarray}
P_2=\frac{q_{\theta0}}{q_{rs0}},
\end{eqnarray}

\begin{eqnarray}
P_3=\frac{q_{r0}}{q_{m0}}+\frac{\Gamma_2}{q_{rs0}},
\end{eqnarray}
and
\begin{eqnarray}
P_4=\frac{1}{q_{rs0}-1+\Gamma_2}.
\end{eqnarray}

Finally, the functions $T_i$ can be written as
\begin{eqnarray}
T_1=\frac{q_{rs1}}{q_{rs0}},
\end{eqnarray}

\begin{eqnarray}
T_2=\frac{q_{m1}}{q_{m0}}+\frac{q_{rs1}}{q_{rs0}},
\end{eqnarray}

\begin{eqnarray}
T_3=\frac{q_{rs1}}{q_{rs0}}-\frac{q_{\theta1}}{q_{\theta0}},
\end{eqnarray}

\begin{eqnarray}
T_4=\frac{q_{r1}}{q_{r0}}-\frac{q_{m1}}{q_{m0}},
\end{eqnarray}

\begin{eqnarray}
T_5=\frac{q_{r1}}{q_{r0}}+\frac{q_{\theta1}}{q_{\theta0}},
\end{eqnarray}

\begin{eqnarray}
T_6=\frac{q_{r1}}{q_{r0}}+\frac{q_{\phi1}}{q_{\phi0}},
\end{eqnarray}

\begin{eqnarray}
T_7=\frac{q_{\theta1}}{q_{\theta0}}+\frac{q_{\phi1}}{q_{\phi0}},
\end{eqnarray}

\begin{eqnarray}
T_8=-\frac{2q_{\theta0}}{q_{m0}^3q_{rs0}^3}\left(q_{m1}^2q_{rs0}^2+q_{m0}q_{m1}q_{rs0}q_{rs1}+q_{m1}^2q_{rs1}^2\right).
\end{eqnarray}

The coefficients $q_{m0}(\tau)$, $q_{m1}(\tau)$, $q_{r0}(\tau)$, $q_{r1}(\tau)$, $q_{\theta0}(\tau)$, $q_{\theta1}(\tau)$, $q_{\phi0}(\tau)$, $q_{\phi1}(\tau)$, $q_{rs0}(\tau)$, and $q_{rs1}(\tau)$ are functions of the time ($\tau$). The shell starts at an equatorial radius $q_{rs0}(0)=r_{s0}(0)$ close to the star. Initially the shell is massless, so
$q_{m0}(0)=q_{m1}(0)=q_{r0}(0)=q_{r1}(0)=q_{\theta0}(0)=q_{\theta1}(0)=q_{\phi0}(0)=q_{\phi1}(0)=q_{rs1}(0)=0$. Thus,  to obtain the ratios between the coefficients $q_{ai}$, one expands them for early times $\tau\ll1$,
\begin{eqnarray}
q_{m0}\approx e_{m0}\tau,
\label{eq:qm0}
\end{eqnarray}

\begin{eqnarray}
q_{m1}\approx e_{m1}\tau,
\label{eq:qm1}
\end{eqnarray}

\begin{eqnarray}
q_{r0}\approx e_{r0}\tau,
\label{eq:qr0}
\end{eqnarray}

\begin{eqnarray}
q_{r1}\approx e_{r1}\tau,
\label{eq:qr1}
\end{eqnarray}

\begin{eqnarray}
q_{\theta0}\approx e_{\theta0}\tau,
\label{eq:qtheta0}
\end{eqnarray}

\begin{eqnarray}
q_{\theta1}\approx e_{\theta1}\tau,
\label{eq:qtheta1}
\end{eqnarray}

\begin{eqnarray}
q_{\phi0}\approx e_{\phi0}\tau,
\label{eq:qphi0}
\end{eqnarray}

\begin{eqnarray}
q_{\phi1}\approx e_{\phi1}\tau,
\label{eq:qphi1}
\end{eqnarray}

\begin{eqnarray}
q_{rs0}\approx r_{s0}(0)+e_{rs0}\tau,
\label{eq:qrs0}
\end{eqnarray}

\noindent and
\begin{eqnarray}
q_{rs1}\approx e_{rs1}\tau.
\label{eq:qrs1}
\end{eqnarray}

Substituting these equations in eqns. (\ref{eq:dqm0}) - (\ref{eq:dqrs1}), one obtains
\begin{eqnarray}
e_{m0}=-\cos\eta\left[\alpha\left(\frac{1+A\sin\eta}{1+A/(2n+1)}\right)\left(\frac{\alpha}{\beta}\Lambda-1\right)-\frac{r_{s0}(0)^{3/2}Q_2}{\gamma_2}\left(\Lambda+Q_1\right)\right],
\label{eq:em0}
\end{eqnarray}

\begin{eqnarray}
e_{r0}=-\cos\eta\left[\beta\left(\frac{1+A\sin\eta}{1+A/(2n+1)}\right)\left(\frac{\alpha}{\beta}\Lambda-1\right)+r_{s0}(0)Q_2\left(\Lambda+Q_1\right)\right],
\label{eq:er0}
\end{eqnarray}

\begin{eqnarray}
e_{\theta0}=r_{s0}(0)Q_2\left(\gamma_1-\sin\eta\right)\left(\Lambda+Q_1\right),
\label{eq:etheta0}
\end{eqnarray}

\begin{eqnarray}
e_{\phi0}=-\frac{r_{s0}(0)\sin\theta_0\gamma_3Q_2\left(\Lambda+Q_1\right)}{\gamma_2},
\label{eq:ephi0}
\end{eqnarray}

\begin{eqnarray}
e_{rs0}=\Lambda,
\label{eq:ers0}
\end{eqnarray}

\begin{eqnarray}
e_{m1}&=&-\frac{r_{s0}(0)e_{m0}^2}{e_{\theta0}^2}\left[\frac{e_{\theta0}}{r_{s0}(0)}\Lambda-\frac{2e_{\theta0}e_{\theta1}+e_{\phi0}^2\tan\eta}{e_{m0}r_{s0}(0)}+\frac{r_{0}^{1/2}Q_2}{2}\Lambda\left(\gamma_1-\sin\eta\right)\left(\frac{\gamma_2}{r_{s0}(0)}+\frac{\gamma_4\sin\eta}{\gamma_1^2\gamma_2}\right)\right] \nonumber\\
&+&\Lambda Q_2\left(\frac{r_{s0}(0)e_{m0}^2}{e_{\theta0}^2}\right)\left(\Lambda+Q_1\right)\left[r_{s0}(0)\gamma_4+Q_2\left(\gamma_1-\sin\eta\right)\left(3\gamma_1^2-1-6r_{s0}(0)\gamma_1\gamma_4\right)\right],
\label{eq:em1}
\end{eqnarray}

\begin{eqnarray}
e_{r1}&=&\frac{r_{s0}(0)e_{m0}}{e_{\theta0}}\left[\frac{e_{\theta0}}{r_{s0}(0)}\Lambda-\frac{e_{\theta0}^2+e_{\phi0}^2}{r_{s0}(0)e_{m0}}+\frac{e_{\theta0}\Lambda}{r_{s0}(0)}\left(\frac{e_{m1}}{e_{m0}}-\frac{e_{\theta1}}{e_{\theta0}}\right)\right] \nonumber \\
&+&\cos\eta Q_2\Lambda\left(\frac{r_{s0}(0)e_{m0}}{e_{\theta0}}\right)\left[Q_2\left(\Lambda+Q_1\right)\left(3\gamma_1^2-1-6r_{s0}(0)\gamma_1\gamma_4\right)-\frac{r_{s0}(0)^{1/2}}{2}\left(\frac{\gamma_2}{r_{s0}(0)}+\frac{\sin\eta\gamma_4}{\gamma_1^2\gamma_2}\right)\right],
\label{eq:er1}
\end{eqnarray}

\begin{eqnarray}
e_{\theta1}&=&\frac{r_{s0}(0)Q_2\Lambda\cos\eta}{2}\left[1+\frac{r_{s0}(0)\gamma_4\sin\eta}{\gamma_1^2\gamma_2^2} \right.\nonumber \\
&-&\left.\left(\Lambda+Q_1\right)\left(\frac{r_{s0}(0)^{3/2}\gamma_4\sin\eta}{\gamma_1^2\gamma_2^2}+\frac{r_{s0}(0)^{1/2}}{\gamma_2}+\frac{2r_{s0}(0)^{1/2}Q_2}{\gamma_2}\left(3\gamma_1^2-1-6r_{s0}(0)\gamma_1\gamma_4\right)\right)\right],
\label{eq:etheta1}
\end{eqnarray}

\begin{eqnarray}
e_{\phi1}&=&\frac{r_{s0}(0)e_{m0}}{e_{\theta0}}\left[\frac{e_{\phi0}}{r_{s0}(0)}\Lambda+\frac{e_{\theta0}e_{\phi0}}{r_{s0}(0)e_{m0}}\left(\frac{e_{m1}}{e_{m0}}-\frac{e_{\theta1}}{e_{\theta0}}\tan\eta\right)-\frac{r_{s0}(0)^{1/2}Q_2\gamma_3\Lambda\sin\theta_0}{2\gamma_2}\left(\frac{\gamma_2}{r_{s0}(0)}+\frac{\gamma_4\sin\eta}{\gamma_1^2\gamma_2}\right)\right]+\frac{r_{s0}(0)e_{m0}}{e_{\theta0}}\nonumber \\
&\times&\left(\Lambda+Q_1\right)\left[\frac{r_{s0}(0)Q_2\Lambda\gamma_4\sin\theta_0\sin\eta}{2\gamma_1^2\gamma_3}+\frac{1}{\gamma_2}\left(Q_2\sin\theta_0\left(3\gamma_1^2-1-6r_{s0}(0)\gamma_1\gamma_4\right)-\frac{r_{s0}(0)\gamma_1\gamma_4}{\sin\theta_0}\right)\right],
\label{eq:ephi1}
\end{eqnarray}

\begin{eqnarray}
e_{rs1}=\Lambda\left(\frac{e_{r1}}{e_{r0}}-\frac{e_{m1}}{e_{m0}}\right),
\label{eq:ers1}
\end{eqnarray}
where
\begin{eqnarray}
\Lambda \equiv \frac{e_{r0}}{e_{m0}}.
\label{eq:Lambda}
\end{eqnarray}

\noindent The functions $Q_1$, $Q_2$, and $\gamma_i$ are given by

\begin{eqnarray}
Q_1=\frac{1}{r_{s0}(0)^{1/2}}\left(1+\frac{\sin\eta}{\cos\theta_0}\right)^{1/2},
\end{eqnarray}

\begin{eqnarray}
Q_2=\frac{1}{r_{s0}(0)-1+3\cos^2\theta_0},
\end{eqnarray}

\begin{eqnarray}
\gamma_1=\cos\theta_0,
\end{eqnarray}

\begin{eqnarray}
\gamma_2=\left(1+\frac{\sin\eta}{\cos\theta_0}\right)^{1/2},
\end{eqnarray}

\begin{eqnarray}
\gamma_3=\left(1-\frac{\sin\eta}{\cos\theta_0}\right)^{1/2},
\end{eqnarray}

\noindent and
\begin{eqnarray}
\gamma_4=\frac{\partial\cos\theta_0}{\partial r_s}.
\end{eqnarray}

\noindent Substituting eqns. (\ref{eq:em0}) and (\ref{eq:er0}) in eq. (\ref{eq:Lambda}), one obtains a quadratic function
for $\Lambda$ 
\begin{eqnarray}
&&\left[\frac{\alpha^2}{\beta}\left(\frac{1+A\sin\eta}{1+A/(2n+1)}\right)-\frac{r_{s0}(0)^{3/2}}{\gamma_2}Q_2\right]\Lambda^2-\left[2\alpha\left(\frac{1+A\sin\eta}{1+A/(2n+1)}\right)+r_{s0}(0)Q_2\left(\frac{r_{s0}(0)^{1/2}}{\gamma_2}Q_1+1\right)\right]\Lambda+\nonumber \\
&&\beta\left[\frac{1+A\sin\eta}{1+A/(2n+1)}\right]-r_{s0}(0)Q_1Q_2=0.
\label{eq:lambda2}
\end{eqnarray}

Solving numerically eqns. (\ref{eq:dqm0}) - (\ref{eq:dqrs1}) with initial conditions given by eqns. (\ref{eq:qm0}) -(\ref{eq:qrs1}) for small $\tau$ 
($\tau=10^{-9}$), one finds the values of the coefficients $q_{m0}$, $q_{m1}$, $q_{r0}$, $q_{r1}$, $q_{\theta0}$, $q_{\theta1}$, $q_{\phi0}$, $q_{\phi1}$, $q_{rs0}$, and $q_{rs1}$. Thus, one obtains BCs at the equator for the mass and the momentum fluxes, and the radius as a 
function of time.



\begin{figure*}[!t]
\centering
\includegraphics[scale=0.5]{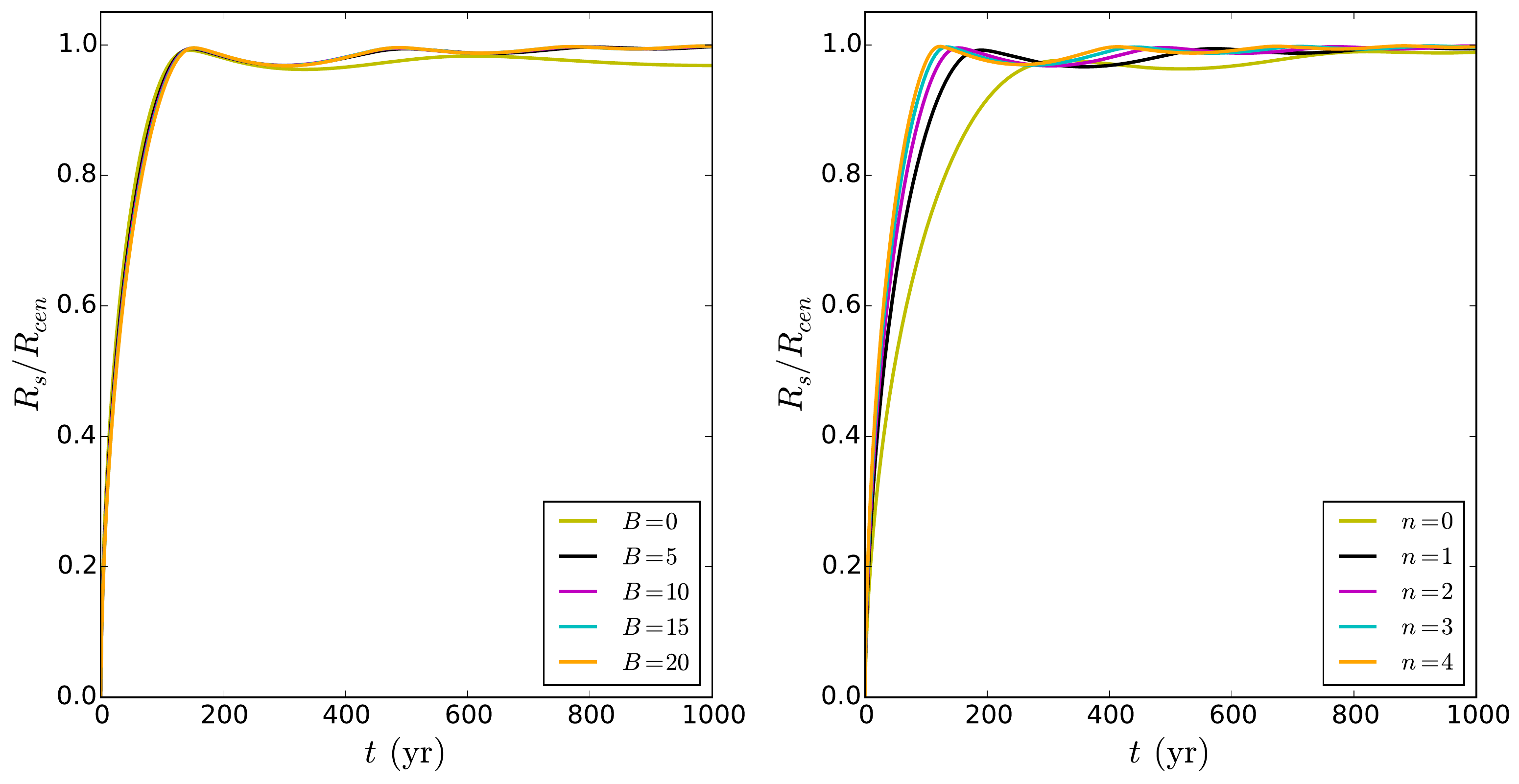}
\caption{Normalized radius at the equator ($\eta=0$) as a function of time for the parameters $\alpha=0.1$, $\beta=21$, and $r_{s0}(0)=10^{-4}$. Left panel: stellar winds with $A=1$, $n=2$, and different values of the anisotropy parameter $B=0$, 5, 10, 15, and 20. Right panel: anisotropic stellar winds with $A=1$, $B=20$, and different exponents $n=0$, 1, 2, 3, and 4.}
\label{fig:requator}
\end{figure*}

\begin{figure*}[!t]
\centering
\includegraphics[scale=0.425]{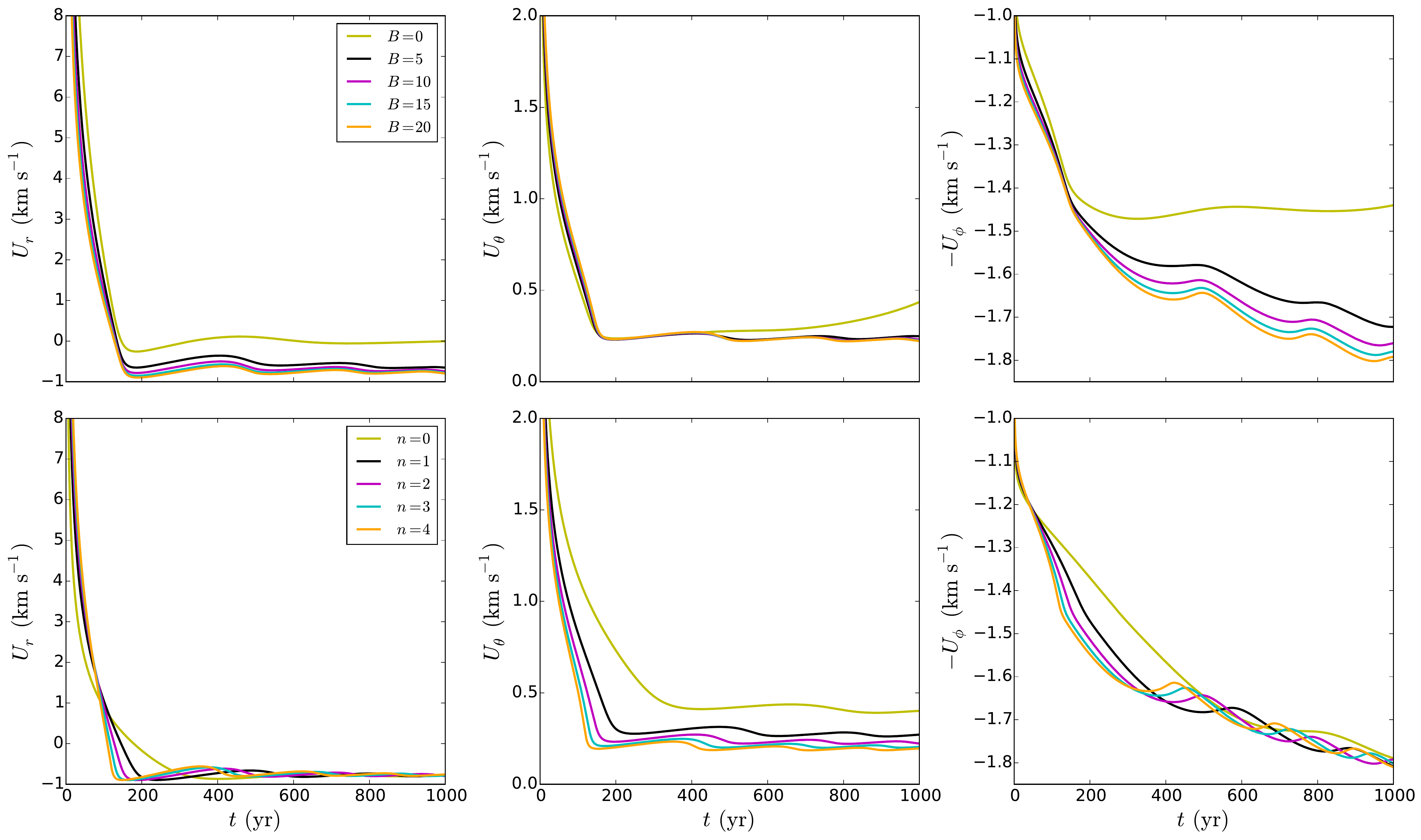}
\caption{Velocity field of the shell at the equator ($\eta=0$) as a function of time for the parameters $\alpha=0.1$, $\beta=21$, and $r_{s0}(0)=10^{-4}$. Upper panels: the radial (left panel), the $\theta$ (middle panel), and the azimuthal (right panel) velocities of the shell for stellar winds with $A=1$, $n=2$, and different values of the anisotropy parameter $B=0$, 5, 10, 15, and 20. Lower panels: the radial (left panel), the $\theta$ (middle panel), and the azimuthal (right panel) velocities of the shell for anisotropic stellar winds with $A=1$, $B=20$, and different exponents $n=0$, 1, 2, 3, and 4.}
\label{fig:velocityeq}
\end{figure*}

Assuming $\alpha=0.1$, $\beta=21$, and $r_{s0}(0) = 10^{-4}$, Figure \ref{fig:requator} shows the evolution of the normalized shell radius for models with parameters $A=1$ and $n=2$, and different values of the anisotropy parameter $B$ (left panel), and models with parameters $A=1$ and $B=20$, and different values of exponent $n$ (right panel). This figure shows that the normalized shell radius grows to a stagnation point at the centrifugal radius. 
For the same models, Figure \ref{fig:velocityeq} shows from left to right, the radial, the $\theta$, and the azimuthal velocities at the equator. 
 With small oscillations, the radial and $\theta$ velocities tend to a constant value, while  the azimuthal velocity decreases with time.

\section{Models of  Wilkin $\&$ Stahler }
\label{app:wilkin}

We compare our models with the outflow models of  \citet{Wilkin_2003} for an isotropic stellar wind ($B=0$). The parameters of the model in their Figure 5 are: a ratio of the wind mass loss rate and the accretion rate $\alpha = 1/3$,  a wind velocity $v_w= 159 {\rm\,km\,s^{-1}} $,
a stellar radius $R_*= 3\,R_{\odot}$, an angular speed of the rotating envelope $\Omega = 2 \times 10^{-14} {\rm\,s^{-1}}$, and a sound speed $a_0 = 0.2 {\rm\,km\,s^{-1}}$.  

We recover the parameters we use in our model in the following way:
from their eq. [1],  the mass accretion rate is $\dot M_a = 1.85 \times 10^{-6}\,M_\odot {\rm\,yr^{-1}}$. 
Thus, the stellar mass, given by their eq. [9], is $M_* = 0.07 \,M_\odot$.
From their eq. [2], the angular and sound speeds give  a centrifugal radius $R_{cen}=0.055$ AU. 
The ratio of the wind and the accretion momentum rates $\beta$ is given by 
$\beta=\alpha v_w/ v_0 = 1.57$, where $v_0 = (G M_*/ R_{\rm cen})^{1/2} = 33.8 {\rm\,km\,s^{-1}}$.
 Finally, the non dimensional time (eq. [\ref{eq:tau}]) is $\Delta \tau = 2.03 (\Delta t / 0.016 {\rm\,yr} )$. We also assume 
a disk  as a boundary condition in the equatorial region with an angle  $\eta=5^\circ$.

\begin{figure}[!t]
\centering
\includegraphics[scale=0.4]{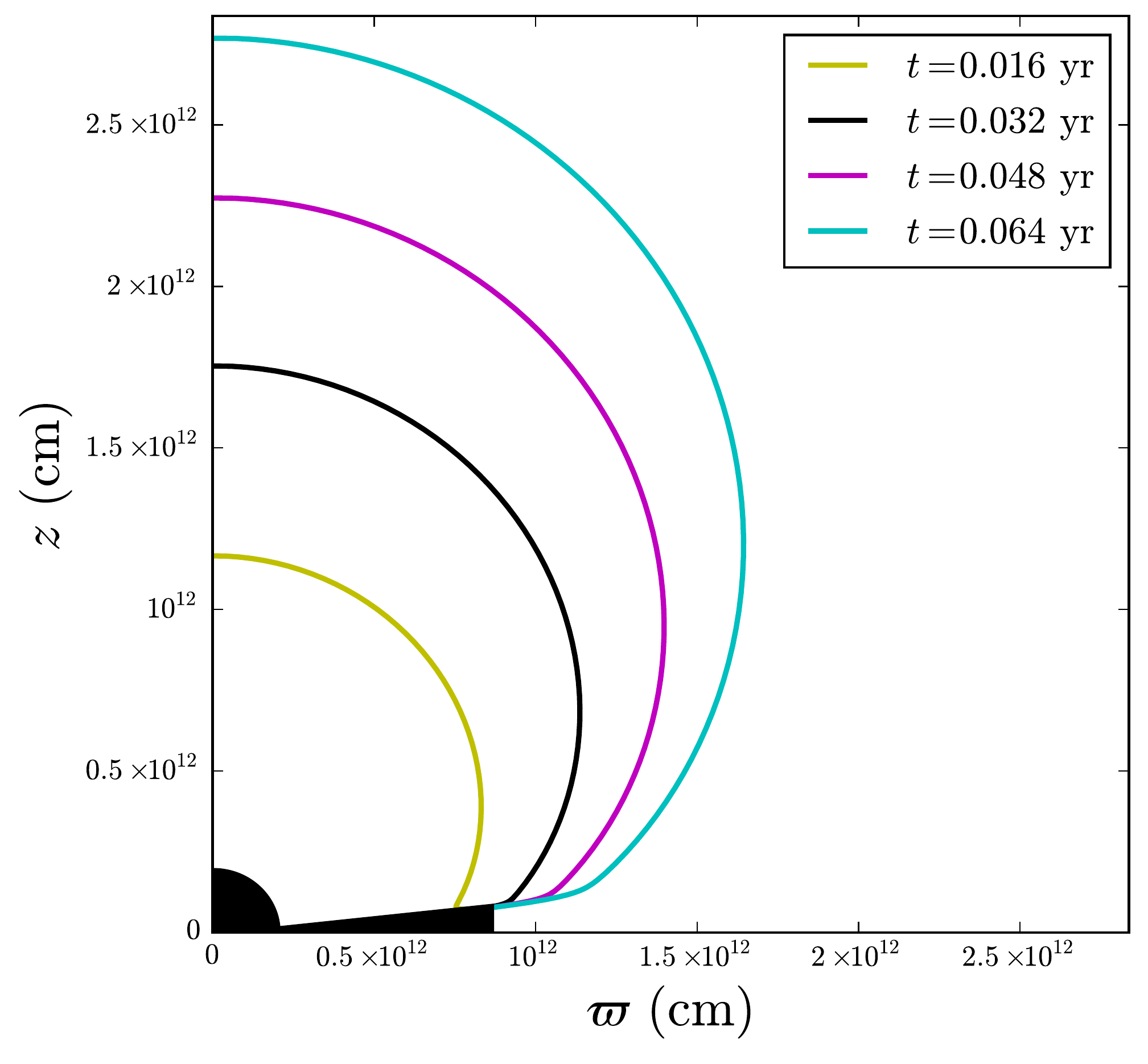}
\caption{Shape of the shell for the same parameters of \citet{Wilkin_2003} for different time steps.}
\label{fig:shellwilkin}
\end{figure}

The results for this integration are plotted in figure \ref{fig:shellwilkin}. The shell  sizes at the pole are the same as the models in 
Figure (5) of Wilkin \& Stahler. Nevertheless, the boundary conditions of both models at the disk surface are different. In our model, the BC is
that the shell radius on the disk surface cannot be larger than the centrifugal radius $R_{\rm cen}$.  In their model, the shell expands continuously beyond the centrifugal radius.

\listofchanges

\end{document}